\documentclass[%
 reprint,
nofootinbib,
 amsmath,amssymb,
 aps,
pra,
floatfix,
]{revtex4-2}
\usepackage{physics}
\usepackage{amsmath}
\usepackage{amsfonts} 
\usepackage{amssymb} 
\usepackage{float}
\usepackage{mathtools}
\usepackage{graphicx}
\usepackage{dcolumn}
\usepackage{bm}
\usepackage{hyperref}
\usepackage{svg}

\begin{document}

\preprint{APS/123-QED}

\title{Preparing spin-squeezed states in Rydberg atom arrays via quantum optimal control}

\author{Edison S. Carrera}
\email{ecarrera@ipht.fr}
\author{Harold Erbin}
\email{harold.erbin@ipht.fr}
\author{Grégoire Misguich}
\email{gregoire.misguich@ipht.fr}
 \affiliation{Université Paris-Saclay, CEA, CNRS, Institut de Physique Théorique, 91191 Gif-sur-Yvette, France}

\date{\today}

\begin{abstract}
We present a quantum optimal control protocol to generate highly spin-squeezed states in Rydberg atom arrays coupled via Ising-type van der Waals interactions. Using gradient-based optimization techniques,  we construct time-dependent pulse sequences that steer an initial product state toward highly entangled, spin-squeezed states with predefined magnetization and squeezing axes.
We focus on the Wineland parameter $\xi_W^2$ to measure spin squeezing, and our approach achieves near-optimal spin squeezing in one-dimensional ring arrays of up to $N=8$ spins, significantly outperforming conventional quench dynamics for all system sizes studied. Remarkably, optimized pulse sequences can be directly scaled to larger arrays without additional optimization, achieving a squeezing parameter as low as $\xi_W^2 = 0.227$ in systems containing $N=50$ spins. This work demonstrates the potential of quantum optimal control methods for preparing highly spin-squeezed states, opening pathways to enhanced quantum metrology.
\end{abstract}

\maketitle


\section{Introduction}

An analog quantum simulator may be defined as a controlled quantum many-body system which can be used to mimic or emulate some other quantum systems of interest~\cite{Feynman1982,cirac_goals_2012,georgescu_quantum_2014,altman_quantum_2021}. This class of systems has seen significant advancements in recent years, and
a large variety of experimental platforms have been developed in this perspective: from trapped ions~\cite{Blatt2012}, superconducting circuits~\cite{Gu2017, Kjaergaard2020}, and ultracold atoms~\cite{Carr2009,gross_quantum_2017,Safronova2018} or atoms in optical tweezers~\cite{bernien_probing_2017,zhang_observation_2017}. They can in principle realize intricate quantum phases of matter and/or dynamical phenomena that are difficult or impossible to study on classical computers. A key feature of quantum simulators is their ability to generate highly entangled states, which play a crucial role in understanding quantum correlations, phase transitions, and nonequilibrium dynamics~\cite{Joshi2023}.\\  

Among various quantum simulation platforms, Rydberg atom arrays have emerged as one of the most powerful tools for engineering exotic many-body states~\cite{Samajdar2020, semeghini_probing_2021,Ebadi2021,EVERED_ProbingKitaevHoneycomb_2025, ott2024probingtopologicalentanglementlarge}.  Their long-range interactions, combined with quantum optimal control techniques~\cite{Koch_2022}—which determine the external fields needed to steer the system toward a desired quantum state—have been employed to realize topological phases~\cite{Perciavalle_2024, Kalinowski_2023}, fully connected cluster states~\cite{Crescimanna_2023}, and optimized quantum gates~\cite{Jandura_2022}.\\

Another area where highly controllable quantum systems prove valuable is quantum metrology. For instance, to measure a magnetic field, one can exploit the precession of an ensemble of $N$ spins-$\frac{1}{2}$ and perform some phase estimation~\cite{Pezz__2018}. This typically involves measuring the magnetization of the spins along a chosen axis after a certain precession time.
The precision of such a measurement is fundamentally limited by statistical fluctuations and by what is often referred to as  quantum projection noise. If the spins are prepared in an separable state, the phase sensitivity $\delta\phi$ is bounded by the so-called standard quantum limit (SQL): $\delta\phi_{\rm SQL}=1/\sqrt{N\nu}$, where $\nu$ is the number of measurements.
However, by preparing the system in some appropriately correlated quantum states, one can reduce the statistical noise and surpass the SQL. A notable example of such states are spin-squeezed states (SSS)~\cite{Ma_2011,Pezz__2018}.
Specifically, a SSS achieves a phase sensitivity,  
\begin{equation}
    \sqrt{\nu}\delta\phi = \dfrac{\Delta \hat{J}_{\vec{m}}}{|\langle \hat{J}_{\vec{n}}\rangle|}<\dfrac{1}{\sqrt{N}},
    \label{eq:phi}
\end{equation}
where $\hat{J}_{\vec\mu} = (1/2)\sum_{j=0}^{N-1} \hat{\sigma}_{j}^{\vec{\mu}}$ is the total magnetization operator
along the direction $\vec\mu$, and $(\Delta \hat{J}_{\vec{m}})^2=\langle \hat{J}_{\vec{m}}^2\rangle - \langle \hat{J}_{\vec{m}} \rangle^2$ quantifies the quantum uncertainty in the total magnetization along a direction $\vec{m}$ orthogonal to the mean spin direction $\vec{n}$.

Several criteria exist to identify spin squeezing, including the widely used \textit{Wineland parameter}~\cite{Wine_PhysRevA.46.R6797,wineland_squeezed_1994}:
\begin{equation}
    \xi_W^{2}= \dfrac{N (\Delta \hat{J}_{\vec{m}})^2}{\langle \hat{J}_{\vec{n}}\rangle^2}.
\label{eq:def_Wineland}
\end{equation}
For a fully polarized product state (and $\vec{n}$ chosen along the polarization axis) we have $\xi_W^{2}=1$. On the other hand, states with $\xi_W^{2} < 1$ are said to be spin squeezed and their associated phase sensitivity can be better than the SQL. 
The Wineland parameter also serves as an entanglement witness~\cite{Duger_PhysRevA.64.052106} in the sense
that a state with $\xi_W^{2} < 1$ cannot be separable. It can also be used to quantify the amount of multipartite entanglement. Indeed, the knowledge of $\xi_W^{2}$ can provide a lower bound on the entanglement depth~\cite{Sorensen_2001}.
$\xi_W^{2}$ is also a very interesting quantity from an experimental point of view since it only involves the total magnetization $\vec J$ and is therefore accessible through global measurements. SSS have been investigated across various quantum platforms, including cold atomic ensembles~\cite{PhysRevLett.102.033601}, trapped ions~\cite{PhysRevLett.86.5870, Franke_2023, Bohnet_2016}, and Bose-Einstein condensates (BECs)~\cite{Strobel_2014}. More recently, Rydberg atom arrays have emerged as a promising platform for realizing SSS~\cite{Bornet_2023, Cheraghi_2022}, leveraging their strong, long-range interactions to generate nonclassical correlations.\\  

Quantum control methods offer powerful tools to manipulate quantum systems using external fields, with the aim of achieving specific objectives~\cite{altafini_modeling_2012,glaserTrainingSchrodingersCat2015,Koch_2022}, such as maximizing fidelity to a desired target state~\cite{doria_optimal_2011,luchnikov_controlling_2024,zengAdiabaticEchoProtocols2025}.
These techniques are particularly useful for generating nonclassical states, including SSS.
For instance, Ref.~\cite{Law_2001} proposes using some tuned external fields combined with (all-to-all) atom-atom interactions to enhance spin squeezing in Bose-Einstein condensates.
Ref.~\cite{shenEfficientSpinSqueezing2013} also focused on Bose-Einstein condensates and studied the generation of SSS with quantum control and all-to-all interactions.
In Ref.~\cite{Pichler_2016} it was shown that optimal control methods can significantly improve spin squeezing even in the presence of noise and imperfections. By dynamically tuning the interaction strength, quantum control enables the preparation of SSS on timescales shorter than those required by adiabatic approaches.
In~\cite{haineMachineDesignedSensorMake2020}, an original method to optimize the sensitivity of quantum sensing using quantum control was proposed.
Techniques based on reinforcement learning~\cite{zhao_strategy_2024} and genetic algorithms~\cite{zhao_preparing_2024} have also been used to find control pulses leading to spin squeezing in spin systems with all-to-all interactions.
Moreover, it has recently been shown~\cite{carrera_testing_2025} that the number of experimental repetitions needed to estimate the Wineland parameter with a statistical error $\epsilon < 1$ scales as $\mathcal{O}(N/(1 -\xi_W^2)^2 \log{1/\epsilon})$.
This means that characterizing SSS experimentally becomes increasingly difficult when the number of spins grows and that it is useful to generate states with $\xi_W^2$ as low as possible.\\

In this work, we propose a quantum optimal control approach to generate highly squeezed states in Rydberg atom arrays with Ising-type van der Waals interactions. A key distinction from several previous studies on the generation of SSS through quantum control lies in the structure of our Hamiltonian, which is dominated by short-range interactions rather than by all-to-all couplings.
Using gradient-based optimization, we construct time-dependent control fields that produce SSS with predefined magnetization and squeezing axes. For systems up to $ N=8 $, our method approaches the lower bound for the Wineland parameter and phase sensitivity. We analyze the unitary dynamics under the optimized pulses and assess the robustness of the protocol in the presence of dephasing noise. Furthermore, we show that pulse sequences optimized for $N=10$ can be seamlessly transferred to larger systems, yielding a Wineland parameter as low as $ \xi_W^2 \approx 0.227 $ for $ N=50 $, compared to \( \xi_W^2 \approx 0.41 \) obtained via conventional quench dynamics.

\begin{figure}
    \centering
    \includegraphics[scale=0.5]{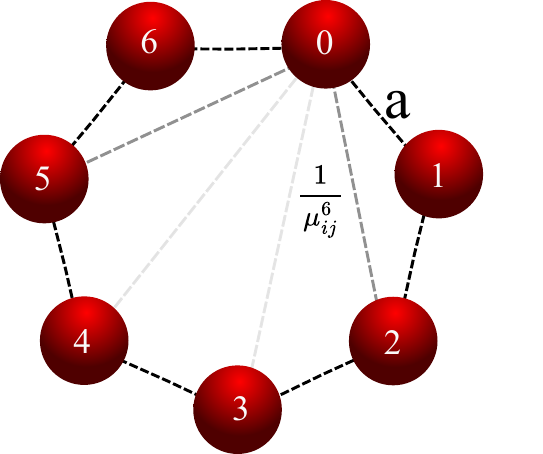}
    \caption{Sketch of the positions of the spins. $a$ denotes the distance between nearest neighbors spins.}
    \label{ring}
\end{figure} 

\section{Model and method}
\subsection{Hamiltonian}

Rydberg atom quantum simulators provide a powerful platform for generating spin-squeezed states. In these setups, individual atoms are confined in optical lattices or optical tweezers and then excited to Rydberg states, where they experience strong and controllable interactions (for a review, see Refs.~\cite{saffman_quantum_2010,browaeys_many-body_2020}). These systems support two main types of interactions: Ising (van der Waals) and XY (dipolar)~\cite{Henriet_2020}. In the present study, we focus on the Ising mode
where the two states of each qubit are the ground state and an excited Rydberg state of the atom. We consider atoms arranged on a ring, as illustrated in Fig.~\ref{ring}, and the interaction Hamiltonian reads
\begin{equation}
    \hat{H}_0 = J \sum_{i\neq j}\frac{1}{\mu_{ij}^{6}} \hat{n}_i \hat{n}_j, 
\end{equation}
where \( \hat{n}_j = (1+\hat{\sigma}_{j}^{z})/2 \) is the Rydberg state occupancy, and \( \mu_{ij} = \|\vec{r_i} - \vec{r_j}\|/a \) is the Euclidean distance between spins \( i \) and \( j \), normalized by the nearest-neighbor spacing \( a \). 
Here, \( J = C_6/a^6 \) represents the interaction energy between nearest neighbors, with \( C_6 \) the van der Waals coefficient that varies rapidly with the principal quantum number of the Rydberg state. In current quantum devices, \( C_6 \) is typically on the order of a few \(10^6\,\text{rad}\cdot\mu\text{m}^6/\mu\text{s}\)~\cite{beguin_direct_2013,wurtz_aquila_2023}, which for typical spacings \( a \sim 10\,\mu\text{m} \) gives \( J \) of the order of a few rad$/\mu$s.

In addition to the Ising interactions $\hat{H}_0$, the Hamiltonian contains a control part $\hat{H}_C$ induced by the laser and which takes the form of an effective magnetic field with transverse ($x$) and longitudinal ($z$) components: 
\begin{equation}
    \hat{H}_C(t) = \Delta(t) \hat{J}_z+ \Omega(t)\hat{J}_x.
\end{equation}

The control fields \(\Delta(t)\) and \(\Omega(t)\), referred to as the \textit{detuning} and \textit{Rabi drive amplitude}, are assumed to be spatially homogeneous and to act uniformly on all atoms (spins).

Ideally, in a noise- and decoherence-free scenario, and without hardware limitations, the physical distance $a$ between atoms would be irrelevant. Indeed, rescaling the nearest-neighbor spacing from \(a\) to \(a' = a/\alpha\) modifies the Ising interaction strength from \(J\) to \(J' = \alpha^6 J\). Then, adjusting the control fields accordingly, \(\Omega(t) \to \alpha^6 \Omega(t)\) and \(\Delta(t) \to \alpha^6 \Delta(t)\), the same final state \(\ket{\psi}\) is reached at a rescaled time \(t'_{\rm final} = \alpha^{-6} t_{\rm final}\).
In practice, however, the maximum control fields achievable by a given device impose a \emph{minimal} interatomic spacing. Also, the motion of the atoms affects the spin dynamics more strongly the closer they are.
Furthermore, competing sources of noise and decoherence limit the timescale over which the dynamics remains approximately unitary, typically to 5--10~\(\mu\)s~\cite{leseleuc_analysis_2018,Henriet_2020,wurtz_aquila_2023}. The distance $a$ must then be sufficiently small for the dynamics arising from the interactions ($\hat H_0$) to be faster than decoherence. So, determining an optimal interatomic distance thus requires quantitative knowledge of both the relevant noise sources and hardware limitations.

A possible approach to generate SSS is to perform a quantum quench, where the qubits are initialized in a simple product state and then evolved with a time-{\em independent} Hamiltonian $\hat{H}=\hat{H}_0+\hat{H}_C$ with fixed $\Delta(t)=\Delta_0$ and $\Omega(t)=\Omega_0$. A significant challenge in experimentally generating a SSS through such a quench is the fact that one does not know in advance when the squeezing (if any) will peak, or the orientation of the mean spin (magnetization) and the squeezing axis. Consequently, the experiment must be repeated multiple times in order to find the optimal SSS. Although experimentally challenging, the authors of Ref.~\cite{Bornet_2023} successfully implemented such a protocol.
By leveraging long-range interactions in a two-dimensional dipolar XY system they obtained SSS with $N=100$ atoms and reached $\xi_W^2 \simeq 0.44$.

We propose employing an optimal-control strategy, using a time-dependent Hamiltonian $\hat{H}(t)=\hat{H}_0+\hat{H}_C(t)$, to steer the initial product state \( \ket{\psi_0} = \ket{\uparrow_z}^{\otimes N} \), which corresponds to the most natural initial state on Rydberg devices, toward a SSS state with a {\em predetermined} squeezing direction, thus
avoiding the need to look for the actual squeezing axis. Moreover, such an optimized time-dependent Hamiltonian can lead to some stronger spin squeezing than a simple quench.

The goal is to optimize the control fields $\Delta(t)$ and $\Omega(t)$ to drive a quantum system from an initial state \(\ket{\psi_0}\) to a final state \(\ket{\psi_T}\) minimizing some cost function \(\mathcal{C}(\ket{\psi_T})\).
The final state is related to the initial state through the following unitary operator:  
\begin{equation}
    \hat{V}(T) = \mathcal{T} \exp\left(-i \int_{0}^{T} \hat{H}(t) dt \right),
\end{equation}
where $\mathcal{T}$ is the time ordering operator and we have set $\hbar=1$.

In the field of optimal quantum control a common choice of cost function is $1-| \braket{\phi}{\psi_T} |^2$ where $\ket{\phi}$ is some target state. We instead build a cost function \(\mathcal{C}\) from the Wineland parameter $\xi_W$ in order to find final states with the largest possible squeezing attainable with the Hamiltonian $\hat{H}_0+\hat{H}_C(t)$.

Quantum optimal control algorithms fall into two main categories: gradient-free~\cite{doria_optimal_2011}, and gradient-based methods. Gradient-based approaches leverage automatic differentiation~\cite{baydin2018automaticdifferentiationmachinelearning} (AD), a computational technique that efficiently evaluates derivatives by systematically applying the chain rule, enabling precise and scalable optimization. We employ here the \textit{Gradient Ascent Pulse Engineering} (GRAPE) method, which utilizes AD to refine control pulses for optimal state preparation~\cite{Khaneja2005, Saywell2018}.
For a recent application of the GRAPE method in a similar many-body context, see \cite{zengAdiabaticEchoProtocols2025}.

\subsection{Piecewise constant control fields}

We parametrize the control fields as piecewise constant functions. In other words, the total evolution time is divided into a finite number of intervals during which both $\Omega$ and $\Delta$ remain constant.
Specifically, we define  
\begin{equation}
    \Delta(t) = u_k^{(z)}, \quad \Omega(t) = u_k^{(x)}, \quad
    \text{for} \quad t \in [t_k, t_{k+1}[
\end{equation}
where $t_k=k\delta t$  for all \(k \in [0,M-1]\), with a total time given by \(T = M \delta t\). 
The time-evolution operator can be expressed as
\begin{equation}
    \hat{V} = \prod_{k=0}^{M-1} \exp\left(-i \delta t \big( \hat{H}_0 + u_{k}^{(x)} \hat{J}_x + u_{k}^{(z)} \hat{J}_z \big)\right),
\end{equation}
where the total time $T$ is fixed and every interval has the same duration $\delta t=t_{k+1}-t_k$.
It is instead advantageous to be able to vary $T$ and to allow for unequal time intervals.
To do so we introduce an additional set of parameters, \(\{ u_{k=0,\cdots,M-1}^{(h)} \}\) as coefficients in front of the interaction Hamiltonian $\hat{H}_0$. The new time evolution operator is then:
\begin{equation}
    \hat{U} = \prod_{k=0}^{M-1} \exp\left(-i \delta t\big( u_k^{(h)} \hat{H}_0 + u_{k}^{(x)} \hat{J}_x + u_{k}^{(z)} \hat{J}_z \big) \right),
    \label{eq:U1}
\end{equation}
which can be rewritten as:
\begin{equation}
    U = \prod_{k=0}^{M-1} \exp\left(-i \delta t_k \big(\hat{H}_0 + \widetilde{u}_{k}^{(x)} \hat{J}_x + \widetilde{u}_{k}^{(z)} \hat{J}_z \big) \right).
    \label{eq:U2}
\end{equation}
Here, the rescaled time interval $\delta t_k =  u_k^{(h)} \delta t$
represents the duration of the control pulse at step \( k \), while 
$\widetilde{u}_{k}^{(\alpha)} = u_{k}^{(\alpha)} / u_k^{(h)}$ denotes the new pulse magnitudes.
To ensure that all time intervals $\delta t_k$ are positive and that the pulse magnitudes $\widetilde{u}_{k}^{(\alpha)}$ remain finite, we fix some constant $\eta>0$ and impose that $u_k^{(h)}>\eta$.
Using the second formulation Eq.~\eqref{eq:U2}, which corresponds to the physical situation where the coefficient in front of the Ising Hamiltonian $\hat{H}_0$ is constant, the total evolution time is now variable and given by 
\begin{equation}
t_{\rm final} = \sum_k \delta t_k > T \eta.
\label{eq:tfinal}
\end{equation}
In the following, the set of data $\left\{ \delta t_k, \tilde u_k^{(x)},\tilde u_k^{(z)}\right\}$ (or equivalently $\left\{ u_k^{(x)},u_k^{(z)},u_k^{(h)}\right\}$) is what we refer to as a pulse sequence, or simply a pulse.  

\begin{figure}
    \centering
    \includegraphics[scale=1]{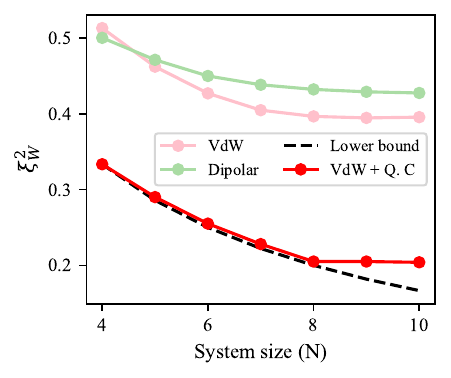}
\caption{\textbf{Summary of the main results.} Optimal Wineland parameter \( \xi_W^2 \) as a function of the system size \( N \). Results are shown for the quench protocol, starting from the initial state \( \ket{\uparrow_y}^{\otimes N} \) under the Ising van der Waals (VdW) and XY dipolar (Dipolar) Hamiltonians, and for the quantum control protocol developed in this work (VdW+QC), starting from \( \ket{\uparrow_z}^{\otimes N} \) with optimized time-dependent fields \( \Omega(t) \) and \( \Delta(t) \). The dashed line indicates the lower bound \( \xi_W^2 = (1+N/2)^{-1} \), corresponding to the minimum achievable value of the Wineland parameter~\cite{Pezz__2018}.}
\label{fig:summary}
\end{figure}

\section{Optimization}
In addition to the Wineland parameter, we wish to control the
direction  of the mean magnetization, as well as the squeezing axis in the final state. In order to do so we define the cost function to be minimized as:
\begin{equation}
    \mathcal{C} = \xi_W^2 + a_4\left(\langle \hat{J}_{y} \rangle^2 + \langle \hat{J}_{x} \rangle^2\right) + \mathcal{G}.
    \label{eq:C}
\end{equation}
In the equation above the second term is added to minimize the projection of the magnetization in the plane orthogonal to the target direction $\vec n=\vec z$. $\mathcal{G}$ is a function added to control the smoothness and duration of the pulse sequence and is discussed in Appendix~\ref{app:opt}.\\

For a given pulse sequence, specified by $3M$ parameters, the final state $\ket{\psi_{t_{\rm final}}}=\hat{U}\ket{\psi_0}$ is computed numerically using Eq.~\eqref{eq:U1}. Taking advantage of the translation invariance of the Hamiltonian and of the initial state one can restrict to the zero-momentum sector and reduce the size of the matrices by a factor $\mathcal{O}(N)$. The matrix exponentials are then computed using the Padé approximation, as implemented in the JAX library~\cite{jax2018github}.\\

The optimization is carried out using the gradient descent method and AD. The details about the optimization algorithm, constraints, and the values of the hyperparameters are discussed in Appendix~\ref{app:opt}.\\

\section{Results}

\subsection{Final squeezed states with $N=8$}

\begin{figure}[h]
    \centering
    \includegraphics[scale=1]{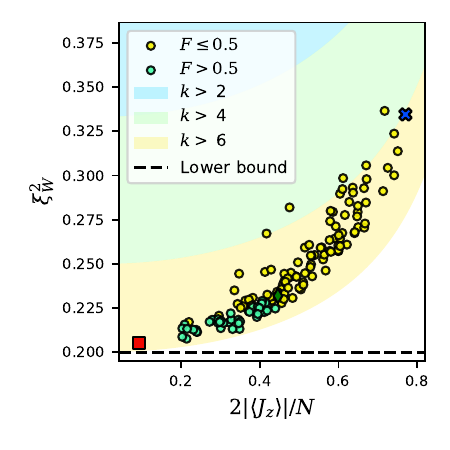}
    \caption{\textbf{Spin-squeezed states obtained from different random initial control pulses for \( N = 8 \).} Each point represents a spin-squeezed state obtained through the application of an optimized control pulse sequence. The vertical axis is the value of Wineland parameter (\(\xi_W^2\)) and the horizontal axis is the normalized mean magnetization (\(2|\langle J_z \rangle|/N\)). The colored regions indicate the depth of entanglement \( k \) according to the criterion developed in \cite{Sorensen_2001}. Yellow points correspond to states with a fidelity $F$ with the GHZ state below 0.5,  and the green points correspond to states with a fidelity larger than 0.5 (see Appendix~\ref{sec:ghz}). 
    The red square marker  highlights the most squeezed state, the green diamond corresponds to the 
    state with the largest magnetization among those with \( F \geq 0.5 \), and the blue cross 
    is for the state with the highest magnetization.}
    \label{fig:histo8}
\end{figure}

\begin{figure*}
    \centering
    \includegraphics[scale=0.25]{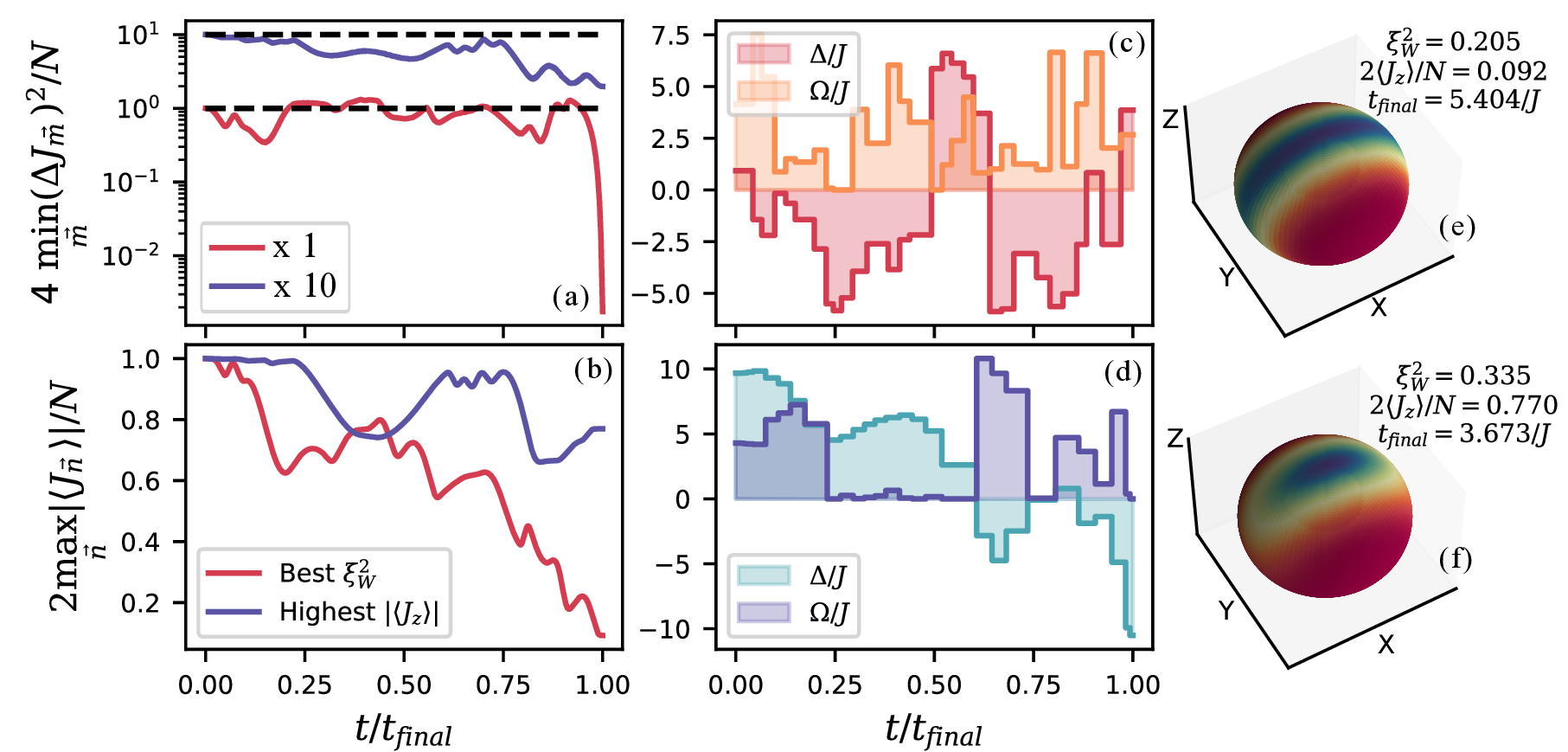}
    \caption{\textbf{Time evolution under two optimized control pulse sequences for \( N = 8 \).} The panels (a) and (b) show the evolution of the minimum variance of the collective angular momentum and the maximum mean magnetization, respectively. For comparison the dashed line represents the variance of a product state (note that the blue curve in (a) has been shifted by a factor 10 for clarity).
    The blue (resp. red) curves correspond to the pulse sequence leading to the state marked with a blue (resp. red) square in 
    Fig.~\ref{fig:histo8}. The associated pulse sequences are presented in (c) and (d). The Q-Husumi function in the final state is displayed in the panels (e) (for the state associated with the red marker) and (f) (state associated with the blue marker). Next to these panels, we show the values of the Wineland parameter, spin length, and the duration (\( t_{\rm final} \)) of the protocols.}
    \label{fig:8dynamics_basic}
\end{figure*}

Optimizing the final state of a many-body system governed by a time-dependent Hamiltonian poses significant numerical challenges, particularly as the system size \( N \) increases---due to the exponential growth of the Hilbert space---or when the number of discrete time steps \( M \) becomes large. We first focus on relatively small spin chains with \( N \leq 10 \) and a moderate number of time steps, \( M = 30 \) (additional results for \( M=100 \) and \( M=3 \) are discussed in Sec.~\ref{sec:M_3_and_100}).  

Figure~\ref{fig:summary} shows the Wineland spin-squeezing parameter achieved by optimized pulses across different system sizes. For systems up to \( N = 8 \), the resulting spin squeezing approaches the absolute lower bound \( \xi_W^2 = 1/(1+N/2) \), which can be saturated by specific linear combinations of three Dicke states~\cite{Pezz__2018}.  

For comparison, we also include the results of a simpler {\em quench} protocol.
There, the system evolves from an initial product state under a time-independent Hamiltonian that includes only spin–spin interactions ($\hat H_0$ in the Ising case), with no external field ($\Delta = \Omega = 0$). We consider both the Ising van der Waals and XY dipolar interactions, relevant to Rydberg atom arrays ~\cite{Henriet_2020}. During the evolution, we determine at each time step the mean spin direction and then search for the optimal squeezing axis in the orthogonal plane. The Wineland parameter is computed accordingly, and the reported results correspond to the minimum value obtained over both time and measurement direction. Notably, the squeezing obtained with optimized pulses significantly exceeds that achieved by the quench protocol in ring geometries of comparable size (see Fig.~\ref{fig:summary}).  
The control landscape is highly non-convex, featuring numerous local minima which complicate the optimization.
On the other hand, the tendency of different random initial conditions to converge to distinct local optima allows us to sample a broad region of the state space. In particular, the states obtained by this method cover a wide range of mean magnetization. It is also possible to add a term in the cost function to target a specific value of the mean magnetization $\langle \hat{J}_z\rangle$ in the final state. Figure ~\ref{fig:histo8} presents the Wineland parameter and the mean magnetization for various output states obtained with \( N = 8 \).

The multipartite entanglement can be evaluated using the Sørensen-Mølmer criterion~\cite{Sorensen_2001}, which relates the value of $\xi_W^2$ to the entanglement depth. While the states considered here are pure and translation invariant—and thus necessarily exhibit genuine multipartite entanglement (GME) unless they are simple product states (which is not the case here)—the criterion remains useful: it provides a lower bound on the entanglement depth for states perturbed infinitesimally away from those considered here. Since these perturbed states can be mixed and/or non translation invariant they are {\it a priori} not guarantied to exhibit any multipartite entanglement. Nevertheless, the Sørensen-Mølmer criterion provides some information on their entanglement depth.
Specifically, if a state  in Fig.~\ref{fig:histo8} lies in a region labeled by $k>p$, all the states in its infinitesimal neighborhood (including mixed ones) possess at least some $p$-partite entanglement. Applying this criterion, we find that nearly all solutions lead to an entanglement depth greater than 6. Furthermore, states with a low Wineland parameter approaching the theoretical lower bound \( 2/(2+N) \), i.e., \( \xi_W^2 \lesssim 0.23 \), have a fidelity $F$ with a Greenberger-Horne-Zeilinger (GHZ) state above 0.5. This implies that the states in their infinitesimal neighborhood exhibit GME (see Appendix~\ref{sec:ghz}).

\subsection{Dynamics with an optimized pulse sequence}

In Fig.~\ref{fig:histo8}, we highlight three particular SSS: the one exhibiting the highest degree of squeezing (red square), a state with both high magnetization and large GHZ fidelity (green diamond), and the state with the highest magnetization among all the states produced (blue cross). The panels (a)-(d) of Fig.~\ref{fig:8dynamics_basic} illustrate the time evolutions leading to the first and third of these final states (highest squeezing and largest magnetization).
Fig.~\ref{fig:8dynamics_basic}(a) shows the time evolution of the magnetization, obtained
by determining at each time step which direction $\vec n(t)$ maximizes $|\langle \hat{J}_{\vec n}\rangle|$.
Panel (b) displays the time evolution of the minimal variance. It is obtained by determining at each time step which direction $\vec m(t)$ orthogonal to $\vec n(t)$ minimizes $(\Delta \hat{J}_{\vec{m}})^2$. The red  (resp. blue) curves correspond to the dynamics leading to the final state marked in red (resp. blue) in Fig.~\ref{fig:histo8}.
Note that due to the choice of the cost function Eq.~\eqref{eq:C}, the mean magnetization in the final state is along the $z$ axis and the variance is minimal in the $y$ direction.
During a large part of the time evolution, the variance --which is directly related to  the correlation between the spins-- fluctuates in some highly nontrivial manner. It is interesting to notice that the variance drops at the end of the protocol. This drop is largely responsible for the low final value of $\xi_W^2$. Such a drop is observed in many cases and it is very pronounced in the case of the pulse leading to the most squeezed state (red curve). The time dependence of the magnetization, displayed in Fig.~\ref{fig:histo8}(b), is no less complex, with some non monotonous behavior. 
The associated pulse sequences are displayed in panels (e) and (f).
Although some parameters favoring some smooth time dependence  of the control fields are present in the cost function (see Appendix~\ref{app:opt}), the optimization process leads to pulses which are rather complex, without any obvious simple pattern. 
It is possible to enforce smoother pulses by increasing the parameters $a_1$ and $a_2$ in the cost function [Eq.~\ref{eq:G} in the Appendix] but we observed that this rapidly deteriorates the value of $\xi_W^2$ in the final state. This indicates that the optimized pulse sequences exploit in a subtle way the interplay between the interaction part of the Hamiltonian and the external control fields along the $z$ and $x$ axis, and some degree of complexity in the time dependence of the control pulses is important to approach highly squeezed states.

To further illustrate the structure of the final states, their Husimi-Q distributions \cite{husimi_formal_1940,Ma_2011} are displayed in the panels (e) and (f) of Fig.~\ref{fig:8dynamics_basic}. 
The Husimi-Q distribution is defined by $f(\vec n)=|\langle \vec n|\psi_{t_{\rm final}}\rangle|^2$
where $|\vec n\rangle$ is a fully polarized product state with magnetization pointing in the $\vec n$ direction. This function allows us to represent graphically  the fluctuations of the direction of the magnetization and to ``visualize'' the associated squeezing. In the case of the state with the strongest squeezing ($\xi_W^2=0.205$) the mean magnetization per site is relatively small ($2\langle \hat{J}_z\rangle/N=0.092$) and we observe an almost complete delocalization of the magnetization direction along an annulus centered around the $y=0$ plane [blue region in Fig.~\ref{fig:8dynamics_basic})(e)]. The squeezing is less extreme for the state with $\xi_W^2=0.335$ (Fig.~\ref{fig:8dynamics_basic} (f)), which has a mean magnetization that is much larger ($2\langle \hat{J}_z\rangle/N=0.770$).

\begin{figure}
    \centering
    \includegraphics[scale=1.]{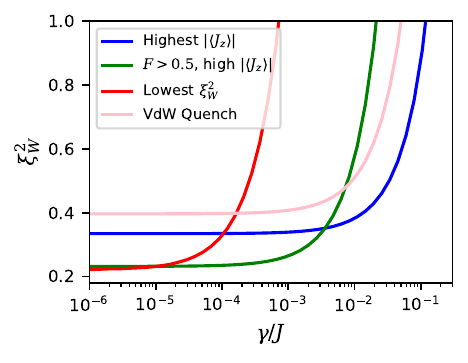}
    \caption{\textbf{Effect of dephasing noise}. Wineland parameter in the final state as a function of the strength $\gamma$ of the dephasing noise (Eq.~\ref{eq:lindblad}) in a system with $N=8$.
    The blue (resp. green and red) curve corresponds to the protocol leading to the state marked in blue (resp. green and red) in Fig.~\ref{fig:histo8}.
    The pink curve corresponds to a quench protocol (time-independent Hamiltonian and no optimization).
    }
    \label{fig:8dephasing}
\end{figure} 

\subsection{Dephasing noise}

Dephasing is an important source of noise in Rydberg atom simulators~\cite{leseleuc_analysis_2018,wurtz_aquila_2023}, and we evaluate here the robustness of the state preparation protocol under this decoherence mechanism. In particular, we consider different pulses (those highlighted in Fig.~\ref{fig:histo8}), where dephasing is added on top of pulses that were optimized \emph{without} decoherence; the problem of optimizing squeezing in the presence of dephasing is left for future work. Using the library DynamiQs~\cite{guilmin2025dynamiqs}, we simulate the Lindblad master equation,
\begin{equation}
\partial_t \hat{\rho} = -i[\hat{H},\hat{\rho}] +\gamma\sum_{m =1}^L \left(\hat{\sigma}^z_m\hat{\rho} \hat{\sigma}^z_m - \hat{\rho}\right),
\label{eq:lindblad}
\end{equation} where $\gamma$ is the dephasing rate.

Figure~\ref{fig:8dephasing} presents the value of Wineland parameter at the final time as a function of $\gamma$. As expected, the Wineland parameter is reduced by dephasing. In addition, we observe that the squeezing is more robust when when the final state has a large magnetization (blue curve) compared to a state with maximal squeezing but lower magnetization (red curve). An intermediate behavior is observed in cases where the squeezing is not optimal but when the final state (in absence of dephasing) has a large overlap with the GHZ state (green curve). Notably, the simple quench protocol is also quite resilient to decoherence (pink curve). But whatever the value of $\gamma$, the  states produced by the time-independent Hamiltonian (quench) remain less squeezed than those obtained with the protocol that produces the SSS with the largest magnetization.

\begin{figure}
    \centering
    \includegraphics[scale=1.1]{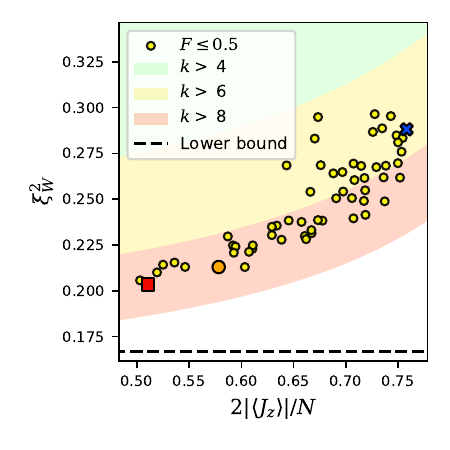}
    \caption{\textbf{Optimized spin-squeezed states obtained from different random initial control fields for \( N = 10 \).} The color code and descriptions are the same as in Fig.~\ref{fig:histo8}. The orange marker refers to the pulse used in larger systems (Figs.~\ref{fig:scaling_wine} and \ref{fig:scaling_corr}).}
    \label{fig:histo10}
\end{figure}

\subsection{Scalability to larger systems}

The computing time required to calculate numerically the final state scales as \( \mathcal{O}\left(M (2^N/N)^3\right) \), and optimizing control pulses for large systems is therefore very challenging. So, to generate 
squeezed states on larger systems, we attempt to repurpose the optimized pulse sequences obtained for small $N$ in larger systems.
We refer to ``scalability'' as the capacity of a pulse sequence to generate squeezed states in systems exceeding the size of the system on which it was initially optimized. For example, most pulses constructed for \( N=8 \) fail to produce high squeezing in larger systems: the lowest Wineland parameter achieved for \( N=12 \) with this method is \( \xi_W^2 \approx 0.29 \), compared to \( \xi_W^2 \approx 0.25 \) in the original \( N=8 \) system. By contrast, some pulses optimized for \( N=10 \) turn out to be more scalable.
 
Figure~\ref{fig:histo10} presents the Wineland parameter and mean magnetization of  various states obtained from different random initial pulses for \( N = 10 \). Using the same criteria used previously to quantify multipartite entanglement, we see that the majority of states are in the $k>8$ region. The highest spin squeezing (state marked with a red square) achieved is \( \xi_W^2 = 0.203 \). Compared to the absolute lower bound ($\xi_W^2 = 1/6$ for $N=10$), this value is not as good as what could be obtained for $N=8$.  But the main advantage of the  pulses  obtained for $N=10$ is that some of them can produce some low value of $\xi_W^2$ on larger systems. In other words, they are scalable.

\begin{figure}
    \centering
    \includegraphics[scale=1]{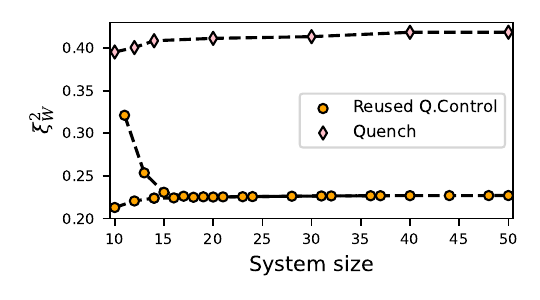}
    \caption{\textbf{Scalability of the optimized control fields for \( N = 10 \).} A pulse optimized for \( N = 10 \) is applied to larger systems. We display the Wineland parameter for both odd and even system sizes, obtained using a 'scalable' optimized pulse (orange square in Fig.~\ref{fig:histo10}). The dynamics under the optimized control fields is computed using the MPS-TDVP2 algorithm with a maximum matrix-product state bond dimension of \( \chi = 300 \) for \textit{Reused Q.control} (tested also with \( \chi = 400 \) ) and \( \chi = 400 \) for \textit{Quench} (tested also with \( \chi = 600 \)). }
    \label{fig:scaling_wine}
\end{figure}

\begin{figure}
    \centering
    \includegraphics[scale=1]{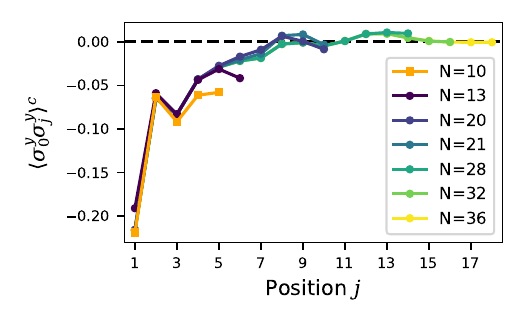}
    \caption{\textbf{Connected two-point spin correlation in the squeezing direction.} A pulse sequence optimized for 10 spins is applied to larger system sizes (\( N \)). We display the connected two-point spin correlation in the \( y \) direction in the final state as a function of the distance $j$ between the two spins. Same simulation parameters as in Fig.~\ref{fig:scaling_wine}.}
    \label{fig:scaling_corr}
\end{figure}

Figure~\ref{fig:scaling_wine} illustrates the scaling of the Wineland parameter
when using a pulse optimized for $N=10$ on larger systems.
For instance, the optimized control pulse that produces \( \xi_W^2 = 0.213 \) for \( N = 10 \) (orange circle in Fig.~\ref{fig:histo10}) generates a SSS with \( \xi_W^2 = 0.222 \) in a system with \( N = 12 \).
The result of such protocol displays some even-odd effect for $N\lesssim 17$ and then the squeezing parameter converges to \( \xi_W^2 \simeq 0.227 \) as \( N \) increases, while the mean magnetization per site $\langle \hat{J}_z\rangle/N$ in the final state remains  nearly constant. Since by definition Var$(\hat{J}_y)=\xi_W^2 \langle J_z\rangle^2/N$, we conclude that variance behaves as Var$(\hat{J}_y) \sim \text{const} \cdot N$. This is compatible with states having nonzero connected correlations only over some finite correlation length that is independent of $N$.
Since the duration of the control pulse is fixed in this setup ($t_{\rm final} = 4.21/J$), correlations can only propagate up to some finite distance, even in large systems.
This is what is observed in Fig.~\ref{fig:scaling_corr}, where correlations appear to be negligible beyond $\sim 10$ lattice spacings.

The spin squeezing in quantum critical models has been studied in \cite{vidalEntanglementSecondorderQuantum2004,frerotQuantumCriticalMetrology2018}. In \cite{frerotQuantumCriticalMetrology2018} the long-distance correlations present in the ground states of Ising models at their quantum critical point were shown to produce SSS. Such a squeezing was also shown to persist at finite temperature in the vicinity of the critical point. For instance, in one space dimension the Wineland parameter in the ground state of an Ising chain with $N=50$ was found to be $\xi^2_W\simeq 0.35$ (-4.5dB) (see Fig.~1 in the Supplemental Material of \cite{frerotQuantumCriticalMetrology2018}). This value is however above the one obtained with our protocol on the same system size. In addition, generating such a ground state would require a very slow adiabatic (or quasi adiabatic) preparation protocol.

It is also important to note that this protocol allows one to prepare states which are significantly more squeezed than those obtained with a simple quench protocol (see diamonds in Fig.~\ref{fig:scaling_wine}). In addition, the protocol with quantum control leads to a faster state preparation. For instance, without quantum control and for $N=50$, the best spin squeezing is observed at $t_{\rm final} = 14.24/J$ (to be compared to $t_{\rm final} = 4.21 /J$ for the optimized pulse). Note also that in the quench protocol the time at which the squeezing is maximum is not known {\it a priori}. So, the data in Fig.~\ref{fig:scaling_wine} has been obtained by computing the squeezing direction and the value of the associated Wineland parameter for each time step along the time evolution, up to $t_{\rm final} = 15/J$.

\section{Conclusion}
We have developed a quantum control framework for the generation of SSS in a Rydberg atom quantum simulator. Our results demonstrate precise control over both the magnetization and the squeezing directions of the final state. For few-body systems ($N\leq 8$), we achieve spin-squeezing values approaching the theoretical lower bound. While the optimal bound is not fully reached for $N=10$, we find that the control pulse sequences optimized at this size can be successfully transferred to larger many-body systems, yielding enhanced spin squeezing in shorter times compared to conventional quantum quench protocols or compared to quasi adiabatic state preparation. Moreover, we observe that control strategies prioritizing high magnetization over strong quantum correlations produce states that are more robust against dephasing noise—an important consideration for experimental implementations. 

Beyond the present study, the quantum control approach developed here can be adapted to generate pulses that would respect the hardware constraints of existing Rydberg quantum simulators, allowing experimental tests on real devices. Furthermore, this framework can be extended to two-dimensional geometries and generalized to optimize other figures of merit associated to interesting entangled states, such as 
topological order~\cite{semeghini_probing_2021,verresen_prediction_2021,samajdar_quantum_2021,zengAdiabaticEchoProtocols2025}, paving the way for a broader class of quantum state engineering protocols in such many-body systems.

\section*{Acknowledgements}
We thank K. Hansenne  and T. Desrousseaux  for several useful  discussions. Y.~Zhang, J.-D.~Bancal, and N.~Sangouard are also acknowledged for some previous collaborations on a closely related topic.

This work is supported by France 2030 under French National Research Agency grant No. ANR-22-QMET-0002 (MetriQs-France program and BACQ project in particular) and by PEPR integrated project EPiQ No. ANR-22-PETQ-0007.

\section*{Data availability}
The data that support the findings of this article are openly available~\cite{edison_github}.

\appendix

\section{Numerical pulse optimization}
\label{app:opt}

\subsection{Cost function and regularizers}
\label{app:opt_reg}
In order to control the shape and duration of the pulses, several regularizers are added to the cost function $\mathcal{C}$, as discussed below. To control the duration of the time evolution and to favor short protocols we introduce a penalty for the quantity
\begin{equation}
    \tau = \sum_k |u_k^{(h)}|
\end{equation}
which is proportional to the total time evolution $t_{\rm final}$ (see Eq.~\eqref{eq:tfinal}).
To favor smooth pulses, we introduce a penalty for the discretized first and second derivatives of the control functions.
This is done by introducing the following quantities:
\begin{equation}
    D_1^{(\alpha)} = \sum_k |d_k^{(\alpha)}|^2
\end{equation}
where $d_k^{(\alpha)}$ is the discrete derivative of the control field $\tilde u_k^{(\alpha)}$:
\begin{equation}
    d_k^{(\alpha)} = \frac{\widetilde{u}_{k+1}^{(\alpha)} -  \widetilde{u}_{k}^{(\alpha)}}{\delta_k}.
\end{equation}
Similarly, to control the (discretized) second derivative of the control fields we introduce
\begin{equation}
    D_2^{(\alpha)} = \sum_k |g_k^{(\alpha)}|^2
\end{equation}
with
\begin{eqnarray}
    g_k^{(\alpha)} &=& \dfrac{d_{k+1}^{(\alpha)}-d_{k}^{(\alpha)}}{\delta_k} \\
    &=& \dfrac{\tilde{u}_{k+2}^{(\alpha)}-\tilde{u}_{k+1}^{(\alpha)}}{\delta_{k+1} \delta_k}  - \frac{\tilde{u}_{k+1}^{(\alpha)}-\tilde{u}_{k}^{(\alpha)}}{(\delta_k)^2}.
\end{eqnarray}
The parameters $D_1^{(\alpha)}$ and $D_2^{(\alpha)}$ (with $\alpha=x$ or $\alpha=z$) and the parameter $\tau$ are included in the cost function with coefficients $a_1$, $a_2$, and $a_3$:
\begin{equation}
    \mathcal{G} = a_1 \left( D_1^{(x)} + D_1^{(z)} \right) + a_2 \left( D_2^{(x)} + D_2^{(z)} \right) + a_3\tau.
    \label{eq:G}
\end{equation}
Choosing large $a_1$ and $a_2$ produces smooth pulses at the expense of larger values of $\xi_W^2$. Similarly, large values of $a_3$ generate shorter pulses but with less squeezing. The data presented in Fig.~\ref{fig:histo8} were obtained with $a_1 =0.06$, $a_2=0.007$ and $a_3=0.005$. These values allow one to reach near optimal squeezing for $N=8$ ($\xi_W^2$ only 2.5\% above the absolute lower bound) and can of course be adjusted to suit the specific problem at hand.

\subsection{Pulse optimization with gradient descent and choice of parameters}

The optimization of the $3M$ parameters is carried out using a gradient descent method.
The code uses AD from JAX~\cite{jax2018github} to compute the gradient of the cost function with respect to the $3M$ parameters. Concerning the gradient descent we use the Adaptive Moment Estimation (ADAM) algorithm~\cite{kingma_adam_2017}.
This algorithm accelerates convergence by incorporating momentum and adaptive learning rates, computing first-order (mean) and second-order (variance) moment estimates of the gradients. It adjusts the learning rate for each parameter individually, making it well suited for nonstationary objectives and sparse gradients. To further improve convergence, we implement a learning rate scheduler that halves the initial learning rate (\(l_r = 0.04\)) if no improvement in the cost function is observed after \(A = 400\) learning steps. The total number of learning steps is capped at \(L_{\text{max}} = 4 \times 10^5\).\\

Additionally, we constrain the parameters to obey
\begin{eqnarray}
        0 < u^{(x)}_{k} &<& u_{\max} \\
        |u^{(z)}_{k}| &<& u_{\max} \\
        u^{(h)}_{k} &>& \eta
\end{eqnarray}
with  \( \eta = 0.1 \) and  \( u_{\max} = 8 \). The remaining parameters were chosen empirically to achieve a compromise between high squeezing, smoothness and short pulse duration : $\widetilde{u}_{max}=u_{max}/\eta = 80$, $\eta = 0.1$, $a_4 = 40/N^2$, $A=400$, $a_1 =0.06$, $a_2=0.007$ and $a_3=0.005$ for $M=30$.\\

\begin{figure}
    \centering
    \includegraphics[scale=1]{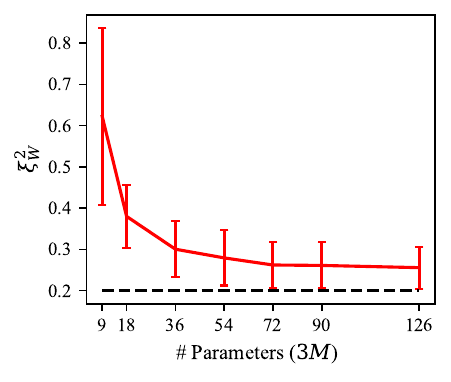}
    \caption{\textbf{Wineland parameter vs number of parameters.} For \( N=8 \), we maximize spin squeezing using pulse sequences with \( M = 3 \) to \( M = 42 \) time steps, starting from 64 different random initial pulse configurations. The upper end of each error bar indicates the maximum value of the Wineland parameter at convergence of the optimization algorithm, while the lower end indicates the minimum. The dashed line shows the theoretical lower bound of the Wineland parameter. All optimizations were performed under the same conditions: the smoothness parameters in Eq.~\eqref{eq:G} were set to \( a_1 = 1.8/M \) and \( a_2 = 0.21/M \), and the duration constraint was fixed to \( a_3 = 0.005 \).
    }   
    \label{fig:n8_paramsextra}
\end{figure}
\section{Variation of the number of steps (\texorpdfstring{$M$}{M})}

\begin{figure}
    \centering
    \includegraphics[scale=1]{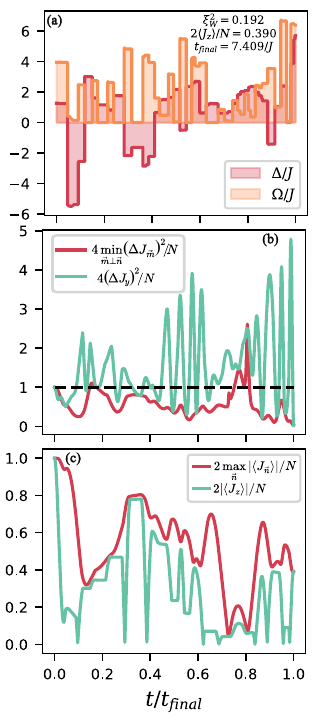}
    \caption{\textbf{Time evolution under optimized pulses for $N = 10$ and $M = 100$.} (a) Optimized control pulses. The Wineland parameter, the magnetization of the output state, and the total duration of the protocol are also indicated.  (b) Time evolution of the variance of the collective spin operator along the direction in which it is minimum. Also, we show the variance in the $y$ direction. (c) Time evolution of the expected magnetization along the direction in which it is maximum, as well as in the $z$ direction. Note that by construction the green and blue curves meet at $t_{\rm final}$ because the cost function requires the final state magnetization to be along $z$ and the minimal variance to be along $y$.}
    \label{fig:n10_100params}
\end{figure}

\subsection{Initialization}

The performance of the optimization algorithm improves when initialized with control pulses that vary slowly in time. Accordingly, we construct the initial control parameters \( u_{k, \text{ini}}^{(\alpha)} \) to be piecewise constant over blocks of \( m \) consecutive time steps. This coarse graining reduces the number of independent parameters defining the initial pulse shape from \( 3M \) to \( 3M/m \), effectively simplifying the initial search space and enhancing convergence. In practice we have used $m=5$.
To initialize the optimization we chose a point in the $3M/m$-dimensional parameter space. It is generated randomly using a truncated uniform distribution with
\( 0 < u_{k, \text{ini}}^{(x)} <  u_{\max} / 2 \), \( |u_{k, \text{ini}}^{(z)}| <  u_{\max} / 2 \), and \( \eta < u_{k, \text{ini}}^{(h)} < 1 \).

\subsection{The importance of the number of parameters}

\begin{figure}
    \centering
    \includegraphics[scale=1]{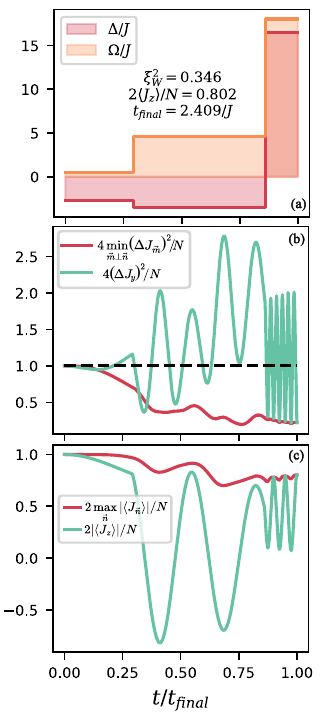}
    \caption{\textbf{Time evolution under optimized pulses for $N = 10$ and $M = 3$.} (a) Optimized control pulses. The Wineland parameter, the final magnetization, and the total duration of the protocol are also indicated.  (b) Time evolution of the variance of the collective spin operator along the direction in which it is minimum. Also, we show the variance in the $y$ direction. (c) Time evolution of the magnetization along the direction in which it is maximum, as well as along the $z$ direction.}
    \label{fig:n10_3params}
\end{figure}

\label{sec:M_3_and_100}

The results presented in the main text were obtained using $M=30$ parameters for each of the three control fields ($x$, $z$ and $h$). Here, we explore the spin squeezing achievable with larger or smaller values of $M$. As previously, we focus on minimizing the Wineland parameter. Multiple parallel optimizations were performed from various initial conditions.

In Fig.~\ref{fig:n8_paramsextra} we show the Wineland parameter for $N=8$ obtained at convergence of the optimization algorithm for \( M = 3 \) to \( M = 42 \). The number of parameters has a clear impact on the minimum value achieved: only for \( M \geq 24 \) does the algorithm reach values relatively close to the Wineland lower bound, \( 2/(2+N) \). However, further increasing the number of parameters does not significantly improve the minimum reachable value. Since the optimization complexity grows linearly with \( M \), it is not advantageous to choose very large values \( M \).

For \( N=10 \) we observe a similar behavior. The optimal pulse yielding the highest spin squeezing for \( M=100 \) is shown in Fig.~\ref{fig:n10_100params}, while the best pulse obtained with \( M=3 \) is displayed in Fig.~\ref{fig:n10_3params}.

For $M=100$, a slight improvement in squeezing is observed relative to $M=30$. However, optimization takes longer and convergence is more challenging due to the increased complexity of the parameter space, resulting in only about $62\%$ of optimization runs converging successfully. Conversely, using fewer parameters ($M=3$) significantly accelerates the optimization process.
Moreover, the simpler parameter space ensures that all optimization runs converge successfully. Also, the relative simplicity of a small-$M$ pulse could make them appropriate for experimental tests.
Nevertheless, the minimal Wineland parameter obtained with $M=3$ ($\xi_W^2=0.346$) remains substantially higher compared to the $M=30$ case ($\xi_W^2=0.206$). We also note that the optimal protocol duration with $M=3$ ($t_{\rm final}=2.41/J$) is shorter than the obtained with $M=100$ ($t_{\rm final}=7.41/J$).
Note, however, that it would also be possible to get shorter pulses for $M=100$ by increasing the parameter $a_3$ in the cost function (Eq.~\ref{eq:G}).\\

\section{GHZ fidelity and genuine multipartite entanglement}
\label{sec:ghz}
\begin{figure}
    \centering
    \includegraphics[scale=0.9]{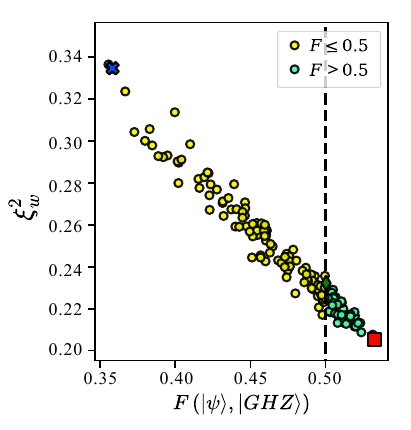}
    \caption{Fidelity with the GHZ state, \(F=|\bra{\psi} {\rm GHZ}\rangle|^2 \), and normalized mean magnetization for the states represented in Fig.~\ref{fig:histo8}.
    }
    \label{fig:GHZ}
\end{figure}
Figure~\ref{fig:GHZ} shows the  fidelity between the states obtained using quantum control (those of Fig.~\ref{fig:histo8}) and the following GHZ state: $1/\sqrt{2} \left( \ket{\uparrow_z^{\otimes N}} + \ket{\downarrow_z^{\otimes N}}\right)$. The spin-squeezing parameter $\xi_W^2$ turns out to be linearly correlated with the fidelity, and the state with the lowest $\xi_W^2$ is the one with the largest overlap with the GHZ state. This observation should be linked to the fact that, from a metrological point of view, a GHZ state saturates the Heisenberg limit for phase sensitivity~\cite{Pezz__2018}. But note that the mean magnetization vanishes in such a state and the Wineland parameter is therefore not defined.

The fidelity with the GHZ state is also an interesting quantity because a state with $F>0.5$ necessarily posses some GME~\cite{toth_detecting_2005, Sackett2000}. Since the states $\ket{\psi}$ produced here are i) pure and ii) translation invariant, we know beforehand they are all GME (unless they are product states, which cannot be the case here since they all display correlations). But the fidelity with the GHZ state nevertheless provides  additional information: any state in the infinitesimal neighborhood of $\ket{\psi}$ must also be GME, even if it is mixed or not translation symmetric. In other words, if a state $\rho$ is obtained by some infinitesimal perturbation of a $\ket{\psi}$ with $F>0.5$, then $\rho$ is also GME. 
We observe that the majority of the strongly squeezed states produced by our protocol indeed have $F>0.5$

\bibliography{thepaper}

\begin{thebibliography}{68}%
\makeatletter
\providecommand \@ifxundefined [1]{%
 \@ifx{#1\undefined}
}%
\providecommand \@ifnum [1]{%
 \ifnum #1\expandafter \@firstoftwo
 \else \expandafter \@secondoftwo
 \fi
}%
\providecommand \@ifx [1]{%
 \ifx #1\expandafter \@firstoftwo
 \else \expandafter \@secondoftwo
 \fi
}%
\providecommand \natexlab [1]{#1}%
\providecommand \enquote  [1]{``#1''}%
\providecommand \bibnamefont  [1]{#1}%
\providecommand \bibfnamefont [1]{#1}%
\providecommand \citenamefont [1]{#1}%
\providecommand \href@noop [0]{\@secondoftwo}%
\providecommand \href [0]{\begingroup \@sanitize@url \@href}%
\providecommand \@href[1]{\@@startlink{#1}\@@href}%
\providecommand \@@href[1]{\endgroup#1\@@endlink}%
\providecommand \@sanitize@url [0]{\catcode `\\12\catcode `\$12\catcode
  `\&12\catcode `\#12\catcode `\^12\catcode `\_12\catcode `\%12\relax}%
\providecommand \@@startlink[1]{}%
\providecommand \@@endlink[0]{}%
\providecommand \url  [0]{\begingroup\@sanitize@url \@url }%
\providecommand \@url [1]{\endgroup\@href {#1}{\urlprefix }}%
\providecommand \urlprefix  [0]{URL }%
\providecommand \Eprint [0]{\href }%
\providecommand \doibase [0]{https://doi.org/}%
\providecommand \selectlanguage [0]{\@gobble}%
\providecommand \bibinfo  [0]{\@secondoftwo}%
\providecommand \bibfield  [0]{\@secondoftwo}%
\providecommand \translation [1]{[#1]}%
\providecommand \BibitemOpen [0]{}%
\providecommand \bibitemStop [0]{}%
\providecommand \bibitemNoStop [0]{.\EOS\space}%
\providecommand \EOS [0]{\spacefactor3000\relax}%
\providecommand \BibitemShut  [1]{\csname bibitem#1\endcsname}%
\let\auto@bib@innerbib\@empty
\bibitem [{\citenamefont {Feynman}(1982)}]{Feynman1982}%
  \BibitemOpen
  \bibfield  {author} {\bibinfo {author} {\bibfnamefont {R.~P.}\ \bibnamefont
  {Feynman}},\ }\bibfield  {title} {\bibinfo {title} {Simulating physics with
  computers},\ }\href {https://doi.org/10.1007/bf02650179} {\bibfield
  {journal} {\bibinfo  {journal} {Int. J. of Theor. Phys.}\ }\textbf {\bibinfo
  {volume} {21}},\ \bibinfo {pages} {467–488} (\bibinfo {year}
  {1982})}\BibitemShut {NoStop}%
\bibitem [{\citenamefont {Cirac}\ and\ \citenamefont
  {Zoller}(2012)}]{cirac_goals_2012}%
  \BibitemOpen
  \bibfield  {author} {\bibinfo {author} {\bibfnamefont {J.~I.}\ \bibnamefont
  {Cirac}}\ and\ \bibinfo {author} {\bibfnamefont {P.}~\bibnamefont {Zoller}},\
  }\bibfield  {title} {\bibinfo {title} {Goals and opportunities in quantum
  simulation},\ }\href {https://doi.org/10.1038/nphys2275} {\bibfield
  {journal} {\bibinfo  {journal} {Nature Phys.}\ }\textbf {\bibinfo {volume}
  {8}},\ \bibinfo {pages} {264} (\bibinfo {year} {2012})}\BibitemShut {NoStop}%
\bibitem [{\citenamefont {Georgescu}\ \emph {et~al.}(2014)\citenamefont
  {Georgescu}, \citenamefont {Ashhab},\ and\ \citenamefont
  {Nori}}]{georgescu_quantum_2014}%
  \BibitemOpen
  \bibfield  {author} {\bibinfo {author} {\bibfnamefont {I.}~\bibnamefont
  {Georgescu}}, \bibinfo {author} {\bibfnamefont {S.}~\bibnamefont {Ashhab}},\
  and\ \bibinfo {author} {\bibfnamefont {F.}~\bibnamefont {Nori}},\ }\bibfield
  {title} {\bibinfo {title} {Quantum simulation},\ }\href
  {https://doi.org/10.1103/RevModPhys.86.153} {\bibfield  {journal} {\bibinfo
  {journal} {Rev. Mod. Phys.}\ }\textbf {\bibinfo {volume} {86}},\ \bibinfo
  {pages} {153} (\bibinfo {year} {2014})}\BibitemShut {NoStop}%
\bibitem [{\citenamefont {Altman}\ \emph {et~al.}(2021)\citenamefont {Altman},
  \citenamefont {Brown}, \citenamefont {Carleo}, \citenamefont {Carr},
  \citenamefont {Demler}, \citenamefont {Chin}, \citenamefont {DeMarco},
  \citenamefont {Economou}, \citenamefont {Eriksson}, \citenamefont {Fu},
  \citenamefont {Greiner}, \citenamefont {Hazzard}, \citenamefont {Hulet},
  \citenamefont {Kollár}, \citenamefont {Lev}, \citenamefont {Lukin},
  \citenamefont {Ma}, \citenamefont {Mi}, \citenamefont {Misra}, \citenamefont
  {Monroe}, \citenamefont {Murch}, \citenamefont {Nazario}, \citenamefont {Ni},
  \citenamefont {Potter}, \citenamefont {Roushan}, \citenamefont {Saffman},
  \citenamefont {Schleier-Smith}, \citenamefont {Siddiqi}, \citenamefont
  {Simmonds}, \citenamefont {Singh}, \citenamefont {Spielman}, \citenamefont
  {Temme}, \citenamefont {Weiss}, \citenamefont {Vučković}, \citenamefont
  {Vuletić}, \citenamefont {Ye},\ and\ \citenamefont
  {Zwierlein}}]{altman_quantum_2021}%
  \BibitemOpen
  \bibfield  {author} {\bibinfo {author} {\bibfnamefont {E.}~\bibnamefont
  {Altman}}, \bibinfo {author} {\bibfnamefont {K.~R.}\ \bibnamefont {Brown}},
  \bibinfo {author} {\bibfnamefont {G.}~\bibnamefont {Carleo}}, \bibinfo
  {author} {\bibfnamefont {L.~D.}\ \bibnamefont {Carr}}, \bibinfo {author}
  {\bibfnamefont {E.}~\bibnamefont {Demler}}, \bibinfo {author} {\bibfnamefont
  {C.}~\bibnamefont {Chin}}, \bibinfo {author} {\bibfnamefont {B.}~\bibnamefont
  {DeMarco}}, \bibinfo {author} {\bibfnamefont {S.~E.}\ \bibnamefont
  {Economou}}, \bibinfo {author} {\bibfnamefont {M.~A.}\ \bibnamefont
  {Eriksson}}, \bibinfo {author} {\bibfnamefont {K.-M.~C.}\ \bibnamefont {Fu}},
  \bibinfo {author} {\bibfnamefont {M.}~\bibnamefont {Greiner}}, \bibinfo
  {author} {\bibfnamefont {K.~R.}\ \bibnamefont {Hazzard}}, \bibinfo {author}
  {\bibfnamefont {R.~G.}\ \bibnamefont {Hulet}}, \bibinfo {author}
  {\bibfnamefont {A.~J.}\ \bibnamefont {Kollár}}, \bibinfo {author}
  {\bibfnamefont {B.~L.}\ \bibnamefont {Lev}}, \bibinfo {author} {\bibfnamefont
  {M.~D.}\ \bibnamefont {Lukin}}, \bibinfo {author} {\bibfnamefont
  {R.}~\bibnamefont {Ma}}, \bibinfo {author} {\bibfnamefont {X.}~\bibnamefont
  {Mi}}, \bibinfo {author} {\bibfnamefont {S.}~\bibnamefont {Misra}}, \bibinfo
  {author} {\bibfnamefont {C.}~\bibnamefont {Monroe}}, \bibinfo {author}
  {\bibfnamefont {K.}~\bibnamefont {Murch}}, \bibinfo {author} {\bibfnamefont
  {Z.}~\bibnamefont {Nazario}}, \bibinfo {author} {\bibfnamefont {K.-K.}\
  \bibnamefont {Ni}}, \bibinfo {author} {\bibfnamefont {A.~C.}\ \bibnamefont
  {Potter}}, \bibinfo {author} {\bibfnamefont {P.}~\bibnamefont {Roushan}},
  \bibinfo {author} {\bibfnamefont {M.}~\bibnamefont {Saffman}}, \bibinfo
  {author} {\bibfnamefont {M.}~\bibnamefont {Schleier-Smith}}, \bibinfo
  {author} {\bibfnamefont {I.}~\bibnamefont {Siddiqi}}, \bibinfo {author}
  {\bibfnamefont {R.}~\bibnamefont {Simmonds}}, \bibinfo {author}
  {\bibfnamefont {M.}~\bibnamefont {Singh}}, \bibinfo {author} {\bibfnamefont
  {I.}~\bibnamefont {Spielman}}, \bibinfo {author} {\bibfnamefont
  {K.}~\bibnamefont {Temme}}, \bibinfo {author} {\bibfnamefont {D.~S.}\
  \bibnamefont {Weiss}}, \bibinfo {author} {\bibfnamefont {J.}~\bibnamefont
  {Vučković}}, \bibinfo {author} {\bibfnamefont {V.}~\bibnamefont
  {Vuletić}}, \bibinfo {author} {\bibfnamefont {J.}~\bibnamefont {Ye}},\ and\
  \bibinfo {author} {\bibfnamefont {M.}~\bibnamefont {Zwierlein}},\ }\bibfield
  {title} {\bibinfo {title} {Quantum {Simulators}: {Architectures} and
  {Opportunities}},\ }\href {https://doi.org/10.1103/PRXQuantum.2.017003}
  {\bibfield  {journal} {\bibinfo  {journal} {PRX Quantum}\ }\textbf {\bibinfo
  {volume} {2}},\ \bibinfo {pages} {017003} (\bibinfo {year}
  {2021})}\BibitemShut {NoStop}%
\bibitem [{\citenamefont {Blatt}\ and\ \citenamefont {Roos}(2012)}]{Blatt2012}%
  \BibitemOpen
  \bibfield  {author} {\bibinfo {author} {\bibfnamefont {R.}~\bibnamefont
  {Blatt}}\ and\ \bibinfo {author} {\bibfnamefont {C.~F.}\ \bibnamefont
  {Roos}},\ }\bibfield  {title} {\bibinfo {title} {Quantum simulations with
  trapped ions},\ }\href {https://doi.org/10.1038/nphys2252} {\bibfield
  {journal} {\bibinfo  {journal} {Nature Phys.}\ }\textbf {\bibinfo {volume}
  {8}},\ \bibinfo {pages} {277–284} (\bibinfo {year} {2012})}\BibitemShut
  {NoStop}%
\bibitem [{\citenamefont {Gu}\ \emph {et~al.}(2017)\citenamefont {Gu},
  \citenamefont {Kockum}, \citenamefont {Miranowicz}, \citenamefont {Liu},\
  and\ \citenamefont {Nori}}]{Gu2017}%
  \BibitemOpen
  \bibfield  {author} {\bibinfo {author} {\bibfnamefont {X.}~\bibnamefont
  {Gu}}, \bibinfo {author} {\bibfnamefont {A.~F.}\ \bibnamefont {Kockum}},
  \bibinfo {author} {\bibfnamefont {A.}~\bibnamefont {Miranowicz}}, \bibinfo
  {author} {\bibfnamefont {Y.-x.}\ \bibnamefont {Liu}},\ and\ \bibinfo {author}
  {\bibfnamefont {F.}~\bibnamefont {Nori}},\ }\bibfield  {title} {\bibinfo
  {title} {Microwave photonics with superconducting quantum circuits},\ }\href
  {https://doi.org/10.1016/j.physrep.2017.10.002} {\bibfield  {journal}
  {\bibinfo  {journal} {Phys. Rep.}\ }\textbf {\bibinfo {volume} {718–719}},\
  \bibinfo {pages} {1–102} (\bibinfo {year} {2017})}\BibitemShut {NoStop}%
\bibitem [{\citenamefont {Kjaergaard}\ \emph {et~al.}(2020)\citenamefont
  {Kjaergaard}, \citenamefont {Schwartz}, \citenamefont {Braum\"{u}ller},
  \citenamefont {Krantz}, \citenamefont {Wang}, \citenamefont {Gustavsson},\
  and\ \citenamefont {Oliver}}]{Kjaergaard2020}%
  \BibitemOpen
  \bibfield  {author} {\bibinfo {author} {\bibfnamefont {M.}~\bibnamefont
  {Kjaergaard}}, \bibinfo {author} {\bibfnamefont {M.~E.}\ \bibnamefont
  {Schwartz}}, \bibinfo {author} {\bibfnamefont {J.}~\bibnamefont
  {Braum\"{u}ller}}, \bibinfo {author} {\bibfnamefont {P.}~\bibnamefont
  {Krantz}}, \bibinfo {author} {\bibfnamefont {J.~I.-J.}\ \bibnamefont {Wang}},
  \bibinfo {author} {\bibfnamefont {S.}~\bibnamefont {Gustavsson}},\ and\
  \bibinfo {author} {\bibfnamefont {W.~D.}\ \bibnamefont {Oliver}},\ }\bibfield
   {title} {\bibinfo {title} {Superconducting qubits: Current state of play},\
  }\href {https://doi.org/10.1146/annurev-conmatphys-031119-050605} {\bibfield
  {journal} {\bibinfo  {journal} {Ann. Rev. of Cond. Matt. Phys.}\ }\textbf
  {\bibinfo {volume} {11}},\ \bibinfo {pages} {369–395} (\bibinfo {year}
  {2020})}\BibitemShut {NoStop}%
\bibitem [{\citenamefont {Carr}\ \emph {et~al.}(2009)\citenamefont {Carr},
  \citenamefont {DeMille}, \citenamefont {Krems},\ and\ \citenamefont
  {Ye}}]{Carr2009}%
  \BibitemOpen
  \bibfield  {author} {\bibinfo {author} {\bibfnamefont {L.~D.}\ \bibnamefont
  {Carr}}, \bibinfo {author} {\bibfnamefont {D.}~\bibnamefont {DeMille}},
  \bibinfo {author} {\bibfnamefont {R.~V.}\ \bibnamefont {Krems}},\ and\
  \bibinfo {author} {\bibfnamefont {J.}~\bibnamefont {Ye}},\ }\bibfield
  {title} {\bibinfo {title} {Cold and ultracold molecules: science, technology
  and applications},\ }\href {https://doi.org/10.1088/1367-2630/11/5/055049}
  {\bibfield  {journal} {\bibinfo  {journal} {New J. Phys.}\ }\textbf {\bibinfo
  {volume} {11}},\ \bibinfo {pages} {055049} (\bibinfo {year}
  {2009})}\BibitemShut {NoStop}%
\bibitem [{\citenamefont {Gross}\ and\ \citenamefont
  {Bloch}(2017)}]{gross_quantum_2017}%
  \BibitemOpen
  \bibfield  {author} {\bibinfo {author} {\bibfnamefont {C.}~\bibnamefont
  {Gross}}\ and\ \bibinfo {author} {\bibfnamefont {I.}~\bibnamefont {Bloch}},\
  }\bibfield  {title} {\bibinfo {title} {Quantum simulations with ultracold
  atoms in optical lattices},\ }\href {https://doi.org/10.1126/science.aal3837}
  {\bibfield  {journal} {\bibinfo  {journal} {Science}\ }\textbf {\bibinfo
  {volume} {357}},\ \bibinfo {pages} {995} (\bibinfo {year}
  {2017})}\BibitemShut {NoStop}%
\bibitem [{\citenamefont {Safronova}\ \emph {et~al.}(2018)\citenamefont
  {Safronova}, \citenamefont {Budker}, \citenamefont {DeMille}, \citenamefont
  {Kimball}, \citenamefont {Derevianko},\ and\ \citenamefont
  {Clark}}]{Safronova2018}%
  \BibitemOpen
  \bibfield  {author} {\bibinfo {author} {\bibfnamefont {M.}~\bibnamefont
  {Safronova}}, \bibinfo {author} {\bibfnamefont {D.}~\bibnamefont {Budker}},
  \bibinfo {author} {\bibfnamefont {D.}~\bibnamefont {DeMille}}, \bibinfo
  {author} {\bibfnamefont {D.~F.~J.}\ \bibnamefont {Kimball}}, \bibinfo
  {author} {\bibfnamefont {A.}~\bibnamefont {Derevianko}},\ and\ \bibinfo
  {author} {\bibfnamefont {C.~W.}\ \bibnamefont {Clark}},\ }\bibfield  {title}
  {\bibinfo {title} {Search for new physics with atoms and molecules},\ }\href
  {https://doi.org/10.1103/revmodphys.90.025008} {\bibfield  {journal}
  {\bibinfo  {journal} {Rev. Mod. Phys.}\ }\textbf {\bibinfo {volume} {90}},\
  \bibinfo {pages} {025008} (\bibinfo {year} {2018})}\BibitemShut {NoStop}%
\bibitem [{\citenamefont {Bernien}\ \emph {et~al.}(2017)\citenamefont
  {Bernien}, \citenamefont {Schwartz}, \citenamefont {Keesling}, \citenamefont
  {Levine}, \citenamefont {Omran}, \citenamefont {Pichler}, \citenamefont
  {Choi}, \citenamefont {Zibrov}, \citenamefont {Endres}, \citenamefont
  {Greiner}, \citenamefont {Vuletić},\ and\ \citenamefont
  {Lukin}}]{bernien_probing_2017}%
  \BibitemOpen
  \bibfield  {author} {\bibinfo {author} {\bibfnamefont {H.}~\bibnamefont
  {Bernien}}, \bibinfo {author} {\bibfnamefont {S.}~\bibnamefont {Schwartz}},
  \bibinfo {author} {\bibfnamefont {A.}~\bibnamefont {Keesling}}, \bibinfo
  {author} {\bibfnamefont {H.}~\bibnamefont {Levine}}, \bibinfo {author}
  {\bibfnamefont {A.}~\bibnamefont {Omran}}, \bibinfo {author} {\bibfnamefont
  {H.}~\bibnamefont {Pichler}}, \bibinfo {author} {\bibfnamefont
  {S.}~\bibnamefont {Choi}}, \bibinfo {author} {\bibfnamefont {A.~S.}\
  \bibnamefont {Zibrov}}, \bibinfo {author} {\bibfnamefont {M.}~\bibnamefont
  {Endres}}, \bibinfo {author} {\bibfnamefont {M.}~\bibnamefont {Greiner}},
  \bibinfo {author} {\bibfnamefont {V.}~\bibnamefont {Vuletić}},\ and\
  \bibinfo {author} {\bibfnamefont {M.~D.}\ \bibnamefont {Lukin}},\ }\bibfield
  {title} {\bibinfo {title} {Probing many-body dynamics on a 51-atom quantum
  simulator},\ }\href {https://doi.org/10.1038/nature24622} {\bibfield
  {journal} {\bibinfo  {journal} {Nature}\ }\textbf {\bibinfo {volume} {551}},\
  \bibinfo {pages} {579} (\bibinfo {year} {2017})}\BibitemShut {NoStop}%
\bibitem [{\citenamefont {Zhang}\ \emph {et~al.}(2017)\citenamefont {Zhang},
  \citenamefont {Pagano}, \citenamefont {Hess}, \citenamefont {Kyprianidis},
  \citenamefont {Becker}, \citenamefont {Kaplan}, \citenamefont {Gorshkov},
  \citenamefont {Gong},\ and\ \citenamefont {Monroe}}]{zhang_observation_2017}%
  \BibitemOpen
  \bibfield  {author} {\bibinfo {author} {\bibfnamefont {J.}~\bibnamefont
  {Zhang}}, \bibinfo {author} {\bibfnamefont {G.}~\bibnamefont {Pagano}},
  \bibinfo {author} {\bibfnamefont {P.~W.}\ \bibnamefont {Hess}}, \bibinfo
  {author} {\bibfnamefont {A.}~\bibnamefont {Kyprianidis}}, \bibinfo {author}
  {\bibfnamefont {P.}~\bibnamefont {Becker}}, \bibinfo {author} {\bibfnamefont
  {H.}~\bibnamefont {Kaplan}}, \bibinfo {author} {\bibfnamefont {A.~V.}\
  \bibnamefont {Gorshkov}}, \bibinfo {author} {\bibfnamefont {Z.-X.}\
  \bibnamefont {Gong}},\ and\ \bibinfo {author} {\bibfnamefont
  {C.}~\bibnamefont {Monroe}},\ }\bibfield  {title} {\bibinfo {title}
  {Observation of a many-body dynamical phase transition with a 53-qubit
  quantum simulator},\ }\href {https://doi.org/10.1038/nature24654} {\bibfield
  {journal} {\bibinfo  {journal} {Nature}\ }\textbf {\bibinfo {volume} {551}},\
  \bibinfo {pages} {601} (\bibinfo {year} {2017})}\BibitemShut {NoStop}%
\bibitem [{\citenamefont {Joshi}\ \emph {et~al.}(2023)\citenamefont {Joshi},
  \citenamefont {Kokail}, \citenamefont {van Bijnen}, \citenamefont {Kranzl},
  \citenamefont {Zache}, \citenamefont {Blatt}, \citenamefont {Roos},\ and\
  \citenamefont {Zoller}}]{Joshi2023}%
  \BibitemOpen
  \bibfield  {author} {\bibinfo {author} {\bibfnamefont {M.~K.}\ \bibnamefont
  {Joshi}}, \bibinfo {author} {\bibfnamefont {C.}~\bibnamefont {Kokail}},
  \bibinfo {author} {\bibfnamefont {R.}~\bibnamefont {van Bijnen}}, \bibinfo
  {author} {\bibfnamefont {F.}~\bibnamefont {Kranzl}}, \bibinfo {author}
  {\bibfnamefont {T.~V.}\ \bibnamefont {Zache}}, \bibinfo {author}
  {\bibfnamefont {R.}~\bibnamefont {Blatt}}, \bibinfo {author} {\bibfnamefont
  {C.~F.}\ \bibnamefont {Roos}},\ and\ \bibinfo {author} {\bibfnamefont
  {P.}~\bibnamefont {Zoller}},\ }\bibfield  {title} {\bibinfo {title}
  {Exploring large-scale entanglement in quantum simulation},\ }\href
  {https://doi.org/10.1038/s41586-023-06768-0} {\bibfield  {journal} {\bibinfo
  {journal} {Nature}\ }\textbf {\bibinfo {volume} {624}},\ \bibinfo {pages}
  {539–544} (\bibinfo {year} {2023})}\BibitemShut {NoStop}%
\bibitem [{\citenamefont {Samajdar}\ \emph {et~al.}(2020)\citenamefont
  {Samajdar}, \citenamefont {Ho}, \citenamefont {Pichler}, \citenamefont
  {Lukin},\ and\ \citenamefont {Sachdev}}]{Samajdar2020}%
  \BibitemOpen
  \bibfield  {author} {\bibinfo {author} {\bibfnamefont {R.}~\bibnamefont
  {Samajdar}}, \bibinfo {author} {\bibfnamefont {W.~W.}\ \bibnamefont {Ho}},
  \bibinfo {author} {\bibfnamefont {H.}~\bibnamefont {Pichler}}, \bibinfo
  {author} {\bibfnamefont {M.~D.}\ \bibnamefont {Lukin}},\ and\ \bibinfo
  {author} {\bibfnamefont {S.}~\bibnamefont {Sachdev}},\ }\bibfield  {title}
  {\bibinfo {title} {Complex density wave orders and quantum phase transitions
  in a model of square-lattice rydberg atom arrays},\ }\href
  {https://doi.org/10.1103/physrevlett.124.103601} {\bibfield  {journal}
  {\bibinfo  {journal} {Phys. Rev. Lett.}\ }\textbf {\bibinfo {volume} {124}},\
  \bibinfo {pages} {103601} (\bibinfo {year} {2020})}\BibitemShut {NoStop}%
\bibitem [{\citenamefont {Semeghini}\ \emph {et~al.}(2021)\citenamefont
  {Semeghini}, \citenamefont {Levine}, \citenamefont {Keesling}, \citenamefont
  {Ebadi}, \citenamefont {Wang}, \citenamefont {Bluvstein}, \citenamefont
  {Verresen}, \citenamefont {Pichler}, \citenamefont {Kalinowski},
  \citenamefont {Samajdar}, \citenamefont {Omran}, \citenamefont {Sachdev},
  \citenamefont {Vishwanath}, \citenamefont {Greiner}, \citenamefont
  {Vuletić},\ and\ \citenamefont {Lukin}}]{semeghini_probing_2021}%
  \BibitemOpen
  \bibfield  {author} {\bibinfo {author} {\bibfnamefont {G.}~\bibnamefont
  {Semeghini}}, \bibinfo {author} {\bibfnamefont {H.}~\bibnamefont {Levine}},
  \bibinfo {author} {\bibfnamefont {A.}~\bibnamefont {Keesling}}, \bibinfo
  {author} {\bibfnamefont {S.}~\bibnamefont {Ebadi}}, \bibinfo {author}
  {\bibfnamefont {T.~T.}\ \bibnamefont {Wang}}, \bibinfo {author}
  {\bibfnamefont {D.}~\bibnamefont {Bluvstein}}, \bibinfo {author}
  {\bibfnamefont {R.}~\bibnamefont {Verresen}}, \bibinfo {author}
  {\bibfnamefont {H.}~\bibnamefont {Pichler}}, \bibinfo {author} {\bibfnamefont
  {M.}~\bibnamefont {Kalinowski}}, \bibinfo {author} {\bibfnamefont
  {R.}~\bibnamefont {Samajdar}}, \bibinfo {author} {\bibfnamefont
  {A.}~\bibnamefont {Omran}}, \bibinfo {author} {\bibfnamefont
  {S.}~\bibnamefont {Sachdev}}, \bibinfo {author} {\bibfnamefont
  {A.}~\bibnamefont {Vishwanath}}, \bibinfo {author} {\bibfnamefont
  {M.}~\bibnamefont {Greiner}}, \bibinfo {author} {\bibfnamefont
  {V.}~\bibnamefont {Vuletić}},\ and\ \bibinfo {author} {\bibfnamefont
  {M.~D.}\ \bibnamefont {Lukin}},\ }\bibfield  {title} {\bibinfo {title}
  {Probing topological spin liquids on a programmable quantum simulator},\
  }\href {https://doi.org/10.1126/science.abi8794} {\bibfield  {journal}
  {\bibinfo  {journal} {Science}\ }\textbf {\bibinfo {volume} {374}},\ \bibinfo
  {pages} {1242} (\bibinfo {year} {2021})}\BibitemShut {NoStop}%
\bibitem [{\citenamefont {Ebadi}\ \emph {et~al.}(2021)\citenamefont {Ebadi},
  \citenamefont {Wang}, \citenamefont {Levine}, \citenamefont {Keesling},
  \citenamefont {Semeghini}, \citenamefont {Omran}, \citenamefont {Bluvstein},
  \citenamefont {Samajdar}, \citenamefont {Pichler}, \citenamefont {Ho},
  \citenamefont {Choi}, \citenamefont {Sachdev}, \citenamefont {Greiner},
  \citenamefont {Vuletić},\ and\ \citenamefont {Lukin}}]{Ebadi2021}%
  \BibitemOpen
  \bibfield  {author} {\bibinfo {author} {\bibfnamefont {S.}~\bibnamefont
  {Ebadi}}, \bibinfo {author} {\bibfnamefont {T.~T.}\ \bibnamefont {Wang}},
  \bibinfo {author} {\bibfnamefont {H.}~\bibnamefont {Levine}}, \bibinfo
  {author} {\bibfnamefont {A.}~\bibnamefont {Keesling}}, \bibinfo {author}
  {\bibfnamefont {G.}~\bibnamefont {Semeghini}}, \bibinfo {author}
  {\bibfnamefont {A.}~\bibnamefont {Omran}}, \bibinfo {author} {\bibfnamefont
  {D.}~\bibnamefont {Bluvstein}}, \bibinfo {author} {\bibfnamefont
  {R.}~\bibnamefont {Samajdar}}, \bibinfo {author} {\bibfnamefont
  {H.}~\bibnamefont {Pichler}}, \bibinfo {author} {\bibfnamefont {W.~W.}\
  \bibnamefont {Ho}}, \bibinfo {author} {\bibfnamefont {S.}~\bibnamefont
  {Choi}}, \bibinfo {author} {\bibfnamefont {S.}~\bibnamefont {Sachdev}},
  \bibinfo {author} {\bibfnamefont {M.}~\bibnamefont {Greiner}}, \bibinfo
  {author} {\bibfnamefont {V.}~\bibnamefont {Vuletić}},\ and\ \bibinfo
  {author} {\bibfnamefont {M.~D.}\ \bibnamefont {Lukin}},\ }\bibfield  {title}
  {\bibinfo {title} {Quantum phases of matter on a 256-atom programmable
  quantum simulator},\ }\href {https://doi.org/10.1038/s41586-021-03582-4}
  {\bibfield  {journal} {\bibinfo  {journal} {Nature}\ }\textbf {\bibinfo
  {volume} {595}},\ \bibinfo {pages} {227–232} (\bibinfo {year}
  {2021})}\BibitemShut {NoStop}%
\bibitem [{\citenamefont {Evered}\ \emph {et~al.}(2025)\citenamefont {Evered},
  \citenamefont {Kalinowski}, \citenamefont {Geim}, \citenamefont {Manovitz},
  \citenamefont {Bluvstein}, \citenamefont {Li}, \citenamefont {Maskara},
  \citenamefont {Zhou}, \citenamefont {Ebadi}, \citenamefont {Xu},
  \citenamefont {Campo}, \citenamefont {Cain}, \citenamefont {Ostermann},
  \citenamefont {Yelin}, \citenamefont {Sachdev}, \citenamefont {Greiner},
  \citenamefont {Vuleti{\'c}},\ and\ \citenamefont
  {Lukin}}]{EVERED_ProbingKitaevHoneycomb_2025}%
  \BibitemOpen
  \bibfield  {author} {\bibinfo {author} {\bibfnamefont {S.~J.}\ \bibnamefont
  {Evered}}, \bibinfo {author} {\bibfnamefont {M.}~\bibnamefont {Kalinowski}},
  \bibinfo {author} {\bibfnamefont {A.~A.}\ \bibnamefont {Geim}}, \bibinfo
  {author} {\bibfnamefont {T.}~\bibnamefont {Manovitz}}, \bibinfo {author}
  {\bibfnamefont {D.}~\bibnamefont {Bluvstein}}, \bibinfo {author}
  {\bibfnamefont {S.~H.}\ \bibnamefont {Li}}, \bibinfo {author} {\bibfnamefont
  {N.}~\bibnamefont {Maskara}}, \bibinfo {author} {\bibfnamefont
  {H.}~\bibnamefont {Zhou}}, \bibinfo {author} {\bibfnamefont {S.}~\bibnamefont
  {Ebadi}}, \bibinfo {author} {\bibfnamefont {M.}~\bibnamefont {Xu}}, \bibinfo
  {author} {\bibfnamefont {J.}~\bibnamefont {Campo}}, \bibinfo {author}
  {\bibfnamefont {M.}~\bibnamefont {Cain}}, \bibinfo {author} {\bibfnamefont
  {S.}~\bibnamefont {Ostermann}}, \bibinfo {author} {\bibfnamefont {S.~F.}\
  \bibnamefont {Yelin}}, \bibinfo {author} {\bibfnamefont {S.}~\bibnamefont
  {Sachdev}}, \bibinfo {author} {\bibfnamefont {M.}~\bibnamefont {Greiner}},
  \bibinfo {author} {\bibfnamefont {V.}~\bibnamefont {Vuleti{\'c}}},\ and\
  \bibinfo {author} {\bibfnamefont {M.~D.}\ \bibnamefont {Lukin}},\ }\bibfield
  {title} {\bibinfo {title} {Probing the {{Kitaev}} honeycomb model on a
  neutral-atom quantum computer},\ }\href
  {https://doi.org/10.1038/s41586-025-09475-0} {\bibfield  {journal} {\bibinfo
  {journal} {Nature}\ }\textbf {\bibinfo {volume} {645}},\ \bibinfo {pages}
  {341} (\bibinfo {year} {2025})}\BibitemShut {NoStop}%
\bibitem [{\citenamefont {Ott}\ \emph {et~al.}(2025)\citenamefont {Ott},
  \citenamefont {Zache}, \citenamefont {Maskara}, \citenamefont {Lukin},
  \citenamefont {Zoller},\ and\ \citenamefont
  {Pichler}}]{ott2024probingtopologicalentanglementlarge}%
  \BibitemOpen
  \bibfield  {author} {\bibinfo {author} {\bibfnamefont {R.}~\bibnamefont
  {Ott}}, \bibinfo {author} {\bibfnamefont {T.~V.}\ \bibnamefont {Zache}},
  \bibinfo {author} {\bibfnamefont {N.}~\bibnamefont {Maskara}}, \bibinfo
  {author} {\bibfnamefont {M.~D.}\ \bibnamefont {Lukin}}, \bibinfo {author}
  {\bibfnamefont {P.}~\bibnamefont {Zoller}},\ and\ \bibinfo {author}
  {\bibfnamefont {H.}~\bibnamefont {Pichler}},\ }\bibfield  {title} {\bibinfo
  {title} {Probing {{Topological Entanglement}} on {{Large Scales}}},\ }\href
  {https://doi.org/10.1103/mdsf-wrbj} {\bibfield  {journal} {\bibinfo
  {journal} {Phys. Rev. Lett.}\ }\textbf {\bibinfo {volume} {135}},\ \bibinfo
  {pages} {090401} (\bibinfo {year} {2025})}\BibitemShut {NoStop}%
\bibitem [{\citenamefont {Koch}\ \emph {et~al.}(2022)\citenamefont {Koch},
  \citenamefont {Boscain}, \citenamefont {Calarco}, \citenamefont {Dirr},
  \citenamefont {Filipp}, \citenamefont {Glaser}, \citenamefont {Kosloff},
  \citenamefont {Montangero}, \citenamefont {Schulte-Herbrüggen},
  \citenamefont {Sugny},\ and\ \citenamefont {Wilhelm}}]{Koch_2022}%
  \BibitemOpen
  \bibfield  {author} {\bibinfo {author} {\bibfnamefont {C.~P.}\ \bibnamefont
  {Koch}}, \bibinfo {author} {\bibfnamefont {U.}~\bibnamefont {Boscain}},
  \bibinfo {author} {\bibfnamefont {T.}~\bibnamefont {Calarco}}, \bibinfo
  {author} {\bibfnamefont {G.}~\bibnamefont {Dirr}}, \bibinfo {author}
  {\bibfnamefont {S.}~\bibnamefont {Filipp}}, \bibinfo {author} {\bibfnamefont
  {S.~J.}\ \bibnamefont {Glaser}}, \bibinfo {author} {\bibfnamefont
  {R.}~\bibnamefont {Kosloff}}, \bibinfo {author} {\bibfnamefont
  {S.}~\bibnamefont {Montangero}}, \bibinfo {author} {\bibfnamefont
  {T.}~\bibnamefont {Schulte-Herbrüggen}}, \bibinfo {author} {\bibfnamefont
  {D.}~\bibnamefont {Sugny}},\ and\ \bibinfo {author} {\bibfnamefont {F.~K.}\
  \bibnamefont {Wilhelm}},\ }\bibfield  {title} {\bibinfo {title} {Quantum
  optimal control in quantum technologies. strategic report on current status,
  visions and goals for research in europe},\ }\href
  {https://doi.org/10.1140/epjqt/s40507-022-00138-x} {\bibfield  {journal}
  {\bibinfo  {journal} {EPJ Quantum Technology}\ }\textbf {\bibinfo {volume}
  {9}},\ \bibinfo {pages} {19} (\bibinfo {year} {2022})}\BibitemShut {NoStop}%
\bibitem [{\citenamefont {Perciavalle}\ \emph {et~al.}(2024)\citenamefont
  {Perciavalle}, \citenamefont {Rossini}, \citenamefont {Polo}, \citenamefont
  {Morsch},\ and\ \citenamefont {Amico}}]{Perciavalle_2024}%
  \BibitemOpen
  \bibfield  {author} {\bibinfo {author} {\bibfnamefont {F.}~\bibnamefont
  {Perciavalle}}, \bibinfo {author} {\bibfnamefont {D.}~\bibnamefont
  {Rossini}}, \bibinfo {author} {\bibfnamefont {J.}~\bibnamefont {Polo}},
  \bibinfo {author} {\bibfnamefont {O.}~\bibnamefont {Morsch}},\ and\ \bibinfo
  {author} {\bibfnamefont {L.}~\bibnamefont {Amico}},\ }\bibfield  {title}
  {\bibinfo {title} {Quantum superpositions of current states in rydberg-atom
  networks},\ }\href {https://doi.org/10.1103/physrevresearch.6.043025}
  {\bibfield  {journal} {\bibinfo  {journal} {Phys. Rev. Research}\ }\textbf
  {\bibinfo {volume} {6}},\ \bibinfo {pages} {043025} (\bibinfo {year}
  {2024})}\BibitemShut {NoStop}%
\bibitem [{\citenamefont {Kalinowski}\ \emph {et~al.}(2023)\citenamefont
  {Kalinowski}, \citenamefont {Maskara},\ and\ \citenamefont
  {Lukin}}]{Kalinowski_2023}%
  \BibitemOpen
  \bibfield  {author} {\bibinfo {author} {\bibfnamefont {M.}~\bibnamefont
  {Kalinowski}}, \bibinfo {author} {\bibfnamefont {N.}~\bibnamefont
  {Maskara}},\ and\ \bibinfo {author} {\bibfnamefont {M.~D.}\ \bibnamefont
  {Lukin}},\ }\bibfield  {title} {\bibinfo {title} {Non-abelian floquet spin
  liquids in a digital rydberg simulator},\ }\href
  {https://doi.org/10.1103/physrevx.13.031008} {\bibfield  {journal} {\bibinfo
  {journal} {Phys. Rev. X}\ }\textbf {\bibinfo {volume} {13}},\ \bibinfo
  {pages} {031008} (\bibinfo {year} {2023})}\BibitemShut {NoStop}%
\bibitem [{\citenamefont {Crescimanna}\ \emph {et~al.}(2023)\citenamefont
  {Crescimanna}, \citenamefont {Taylor}, \citenamefont {Goldberg},\ and\
  \citenamefont {Heshami}}]{Crescimanna_2023}%
  \BibitemOpen
  \bibfield  {author} {\bibinfo {author} {\bibfnamefont {V.}~\bibnamefont
  {Crescimanna}}, \bibinfo {author} {\bibfnamefont {J.}~\bibnamefont {Taylor}},
  \bibinfo {author} {\bibfnamefont {A.~Z.}\ \bibnamefont {Goldberg}},\ and\
  \bibinfo {author} {\bibfnamefont {K.}~\bibnamefont {Heshami}},\ }\bibfield
  {title} {\bibinfo {title} {Quantum control of rydberg atoms for mesoscopic
  quantum state and circuit preparation},\ }\href
  {https://doi.org/10.1103/physrevapplied.20.034019} {\bibfield  {journal}
  {\bibinfo  {journal} {Phys. Rev. Appl.}\ }\textbf {\bibinfo {volume} {20}},\
  \bibinfo {pages} {034019} (\bibinfo {year} {2023})}\BibitemShut {NoStop}%
\bibitem [{\citenamefont {Jandura}\ and\ \citenamefont
  {Pupillo}(2022)}]{Jandura_2022}%
  \BibitemOpen
  \bibfield  {author} {\bibinfo {author} {\bibfnamefont {S.}~\bibnamefont
  {Jandura}}\ and\ \bibinfo {author} {\bibfnamefont {G.}~\bibnamefont
  {Pupillo}},\ }\bibfield  {title} {\bibinfo {title} {Time-optimal two- and
  three-qubit gates for rydberg atoms},\ }\href
  {https://doi.org/10.22331/q-2022-05-13-712} {\bibfield  {journal} {\bibinfo
  {journal} {Quantum}\ }\textbf {\bibinfo {volume} {6}},\ \bibinfo {pages}
  {712} (\bibinfo {year} {2022})}\BibitemShut {NoStop}%
\bibitem [{\citenamefont {Pezzè}\ \emph {et~al.}(2018)\citenamefont {Pezzè},
  \citenamefont {Smerzi}, \citenamefont {Oberthaler}, \citenamefont {Schmied},\
  and\ \citenamefont {Treutlein}}]{Pezz__2018}%
  \BibitemOpen
  \bibfield  {author} {\bibinfo {author} {\bibfnamefont {L.}~\bibnamefont
  {Pezzè}}, \bibinfo {author} {\bibfnamefont {A.}~\bibnamefont {Smerzi}},
  \bibinfo {author} {\bibfnamefont {M.~K.}\ \bibnamefont {Oberthaler}},
  \bibinfo {author} {\bibfnamefont {R.}~\bibnamefont {Schmied}},\ and\ \bibinfo
  {author} {\bibfnamefont {P.}~\bibnamefont {Treutlein}},\ }\bibfield  {title}
  {\bibinfo {title} {Quantum metrology with nonclassical states of atomic
  ensembles},\ }\href {https://doi.org/10.1103/revmodphys.90.035005} {\bibfield
   {journal} {\bibinfo  {journal} {Rev. Mod. Phys.}\ }\textbf {\bibinfo
  {volume} {90}},\ \bibinfo {pages} {035005} (\bibinfo {year}
  {2018})}\BibitemShut {NoStop}%
\bibitem [{\citenamefont {Ma}\ \emph {et~al.}(2011)\citenamefont {Ma},
  \citenamefont {Wang}, \citenamefont {Sun},\ and\ \citenamefont
  {Nori}}]{Ma_2011}%
  \BibitemOpen
  \bibfield  {author} {\bibinfo {author} {\bibfnamefont {J.}~\bibnamefont
  {Ma}}, \bibinfo {author} {\bibfnamefont {X.}~\bibnamefont {Wang}}, \bibinfo
  {author} {\bibfnamefont {C.}~\bibnamefont {Sun}},\ and\ \bibinfo {author}
  {\bibfnamefont {F.}~\bibnamefont {Nori}},\ }\bibfield  {title} {\bibinfo
  {title} {Quantum spin squeezing},\ }\href
  {https://doi.org/10.1016/j.physrep.2011.08.003} {\bibfield  {journal}
  {\bibinfo  {journal} {Phys. Rep.}\ }\textbf {\bibinfo {volume} {509}},\
  \bibinfo {pages} {89–165} (\bibinfo {year} {2011})}\BibitemShut {NoStop}%
\bibitem [{\citenamefont {Wineland}\ \emph {et~al.}(1992)\citenamefont
  {Wineland}, \citenamefont {Bollinger}, \citenamefont {Itano}, \citenamefont
  {Moore},\ and\ \citenamefont {Heinzen}}]{Wine_PhysRevA.46.R6797}%
  \BibitemOpen
  \bibfield  {author} {\bibinfo {author} {\bibfnamefont {D.~J.}\ \bibnamefont
  {Wineland}}, \bibinfo {author} {\bibfnamefont {J.~J.}\ \bibnamefont
  {Bollinger}}, \bibinfo {author} {\bibfnamefont {W.~M.}\ \bibnamefont
  {Itano}}, \bibinfo {author} {\bibfnamefont {F.~L.}\ \bibnamefont {Moore}},\
  and\ \bibinfo {author} {\bibfnamefont {D.~J.}\ \bibnamefont {Heinzen}},\
  }\bibfield  {title} {\bibinfo {title} {Spin squeezing and reduced quantum
  noise in spectroscopy},\ }\href {https://doi.org/10.1103/PhysRevA.46.R6797}
  {\bibfield  {journal} {\bibinfo  {journal} {Phys. Rev. A}\ }\textbf {\bibinfo
  {volume} {46}},\ \bibinfo {pages} {R6797} (\bibinfo {year}
  {1992})}\BibitemShut {NoStop}%
\bibitem [{\citenamefont {Wineland}\ \emph {et~al.}(1994)\citenamefont
  {Wineland}, \citenamefont {Bollinger}, \citenamefont {Itano},\ and\
  \citenamefont {Heinzen}}]{wineland_squeezed_1994}%
  \BibitemOpen
  \bibfield  {author} {\bibinfo {author} {\bibfnamefont {D.~J.}\ \bibnamefont
  {Wineland}}, \bibinfo {author} {\bibfnamefont {J.~J.}\ \bibnamefont
  {Bollinger}}, \bibinfo {author} {\bibfnamefont {W.~M.}\ \bibnamefont
  {Itano}},\ and\ \bibinfo {author} {\bibfnamefont {D.~J.}\ \bibnamefont
  {Heinzen}},\ }\bibfield  {title} {\bibinfo {title} {Squeezed atomic states
  and projection noise in spectroscopy},\ }\href
  {https://doi.org/10.1103/PhysRevA.50.67} {\bibfield  {journal} {\bibinfo
  {journal} {Phys. Rev. A}\ }\textbf {\bibinfo {volume} {50}},\ \bibinfo
  {pages} {67} (\bibinfo {year} {1994})}\BibitemShut {NoStop}%
\bibitem [{\citenamefont {Ulam-Orgikh}\ and\ \citenamefont
  {Kitagawa}(2001)}]{Duger_PhysRevA.64.052106}%
  \BibitemOpen
  \bibfield  {author} {\bibinfo {author} {\bibfnamefont {D.}~\bibnamefont
  {Ulam-Orgikh}}\ and\ \bibinfo {author} {\bibfnamefont {M.}~\bibnamefont
  {Kitagawa}},\ }\bibfield  {title} {\bibinfo {title} {Spin squeezing and
  decoherence limit in ramsey spectroscopy},\ }\href
  {https://doi.org/10.1103/PhysRevA.64.052106} {\bibfield  {journal} {\bibinfo
  {journal} {Phys. Rev. A}\ }\textbf {\bibinfo {volume} {64}},\ \bibinfo
  {pages} {052106} (\bibinfo {year} {2001})}\BibitemShut {NoStop}%
\bibitem [{\citenamefont {Sørensen}\ and\ \citenamefont
  {Mølmer}(2001)}]{Sorensen_2001}%
  \BibitemOpen
  \bibfield  {author} {\bibinfo {author} {\bibfnamefont {A.~S.}\ \bibnamefont
  {Sørensen}}\ and\ \bibinfo {author} {\bibfnamefont {K.}~\bibnamefont
  {Mølmer}},\ }\bibfield  {title} {\bibinfo {title} {Entanglement and extreme
  spin squeezing},\ }\href {https://doi.org/10.1103/physrevlett.86.4431}
  {\bibfield  {journal} {\bibinfo  {journal} {Phys. Rev. Lett.}\ }\textbf
  {\bibinfo {volume} {86}},\ \bibinfo {pages} {4431–4434} (\bibinfo {year}
  {2001})}\BibitemShut {NoStop}%
\bibitem [{\citenamefont {Takano}\ \emph {et~al.}(2009)\citenamefont {Takano},
  \citenamefont {Fuyama}, \citenamefont {Namiki},\ and\ \citenamefont
  {Takahashi}}]{PhysRevLett.102.033601}%
  \BibitemOpen
  \bibfield  {author} {\bibinfo {author} {\bibfnamefont {T.}~\bibnamefont
  {Takano}}, \bibinfo {author} {\bibfnamefont {M.}~\bibnamefont {Fuyama}},
  \bibinfo {author} {\bibfnamefont {R.}~\bibnamefont {Namiki}},\ and\ \bibinfo
  {author} {\bibfnamefont {Y.}~\bibnamefont {Takahashi}},\ }\bibfield  {title}
  {\bibinfo {title} {Spin squeezing of a cold atomic ensemble with the nuclear
  spin of one-half},\ }\href {https://doi.org/10.1103/PhysRevLett.102.033601}
  {\bibfield  {journal} {\bibinfo  {journal} {Phys. Rev. Lett.}\ }\textbf
  {\bibinfo {volume} {102}},\ \bibinfo {pages} {033601} (\bibinfo {year}
  {2009})}\BibitemShut {NoStop}%
\bibitem [{\citenamefont {Meyer}\ \emph {et~al.}(2001)\citenamefont {Meyer},
  \citenamefont {Rowe}, \citenamefont {Kielpinski}, \citenamefont {Sackett},
  \citenamefont {Itano}, \citenamefont {Monroe},\ and\ \citenamefont
  {Wineland}}]{PhysRevLett.86.5870}%
  \BibitemOpen
  \bibfield  {author} {\bibinfo {author} {\bibfnamefont {V.}~\bibnamefont
  {Meyer}}, \bibinfo {author} {\bibfnamefont {M.~A.}\ \bibnamefont {Rowe}},
  \bibinfo {author} {\bibfnamefont {D.}~\bibnamefont {Kielpinski}}, \bibinfo
  {author} {\bibfnamefont {C.~A.}\ \bibnamefont {Sackett}}, \bibinfo {author}
  {\bibfnamefont {W.~M.}\ \bibnamefont {Itano}}, \bibinfo {author}
  {\bibfnamefont {C.}~\bibnamefont {Monroe}},\ and\ \bibinfo {author}
  {\bibfnamefont {D.~J.}\ \bibnamefont {Wineland}},\ }\bibfield  {title}
  {\bibinfo {title} {Experimental demonstration of entanglement-enhanced
  rotation angle estimation using trapped ions},\ }\href
  {https://doi.org/10.1103/PhysRevLett.86.5870} {\bibfield  {journal} {\bibinfo
   {journal} {Phys. Rev. Lett.}\ }\textbf {\bibinfo {volume} {86}},\ \bibinfo
  {pages} {5870} (\bibinfo {year} {2001})}\BibitemShut {NoStop}%
\bibitem [{\citenamefont {Franke}\ \emph {et~al.}(2023)\citenamefont {Franke},
  \citenamefont {Muleady}, \citenamefont {Kaubruegger}, \citenamefont {Kranzl},
  \citenamefont {Blatt}, \citenamefont {Rey}, \citenamefont {Joshi},\ and\
  \citenamefont {Roos}}]{Franke_2023}%
  \BibitemOpen
  \bibfield  {author} {\bibinfo {author} {\bibfnamefont {J.}~\bibnamefont
  {Franke}}, \bibinfo {author} {\bibfnamefont {S.~R.}\ \bibnamefont {Muleady}},
  \bibinfo {author} {\bibfnamefont {R.}~\bibnamefont {Kaubruegger}}, \bibinfo
  {author} {\bibfnamefont {F.}~\bibnamefont {Kranzl}}, \bibinfo {author}
  {\bibfnamefont {R.}~\bibnamefont {Blatt}}, \bibinfo {author} {\bibfnamefont
  {A.~M.}\ \bibnamefont {Rey}}, \bibinfo {author} {\bibfnamefont {M.~K.}\
  \bibnamefont {Joshi}},\ and\ \bibinfo {author} {\bibfnamefont {C.~F.}\
  \bibnamefont {Roos}},\ }\bibfield  {title} {\bibinfo {title}
  {Quantum-enhanced sensing on optical transitions through finite-range
  interactions},\ }\href {https://doi.org/10.1038/s41586-023-06472-z}
  {\bibfield  {journal} {\bibinfo  {journal} {Nature}\ }\textbf {\bibinfo
  {volume} {621}},\ \bibinfo {pages} {740–745} (\bibinfo {year}
  {2023})}\BibitemShut {NoStop}%
\bibitem [{\citenamefont {Bohnet}\ \emph {et~al.}(2016)\citenamefont {Bohnet},
  \citenamefont {Sawyer}, \citenamefont {Britton}, \citenamefont {Wall},
  \citenamefont {Rey}, \citenamefont {Foss-Feig},\ and\ \citenamefont
  {Bollinger}}]{Bohnet_2016}%
  \BibitemOpen
  \bibfield  {author} {\bibinfo {author} {\bibfnamefont {J.~G.}\ \bibnamefont
  {Bohnet}}, \bibinfo {author} {\bibfnamefont {B.~C.}\ \bibnamefont {Sawyer}},
  \bibinfo {author} {\bibfnamefont {J.~W.}\ \bibnamefont {Britton}}, \bibinfo
  {author} {\bibfnamefont {M.~L.}\ \bibnamefont {Wall}}, \bibinfo {author}
  {\bibfnamefont {A.~M.}\ \bibnamefont {Rey}}, \bibinfo {author} {\bibfnamefont
  {M.}~\bibnamefont {Foss-Feig}},\ and\ \bibinfo {author} {\bibfnamefont
  {J.~J.}\ \bibnamefont {Bollinger}},\ }\bibfield  {title} {\bibinfo {title}
  {Quantum spin dynamics and entanglement generation with hundreds of trapped
  ions},\ }\href {https://doi.org/10.1126/science.aad9958} {\bibfield
  {journal} {\bibinfo  {journal} {Science}\ }\textbf {\bibinfo {volume}
  {352}},\ \bibinfo {pages} {1297–1301} (\bibinfo {year} {2016})}\BibitemShut
  {NoStop}%
\bibitem [{\citenamefont {Strobel}\ \emph {et~al.}(2014)\citenamefont
  {Strobel}, \citenamefont {Muessel}, \citenamefont {Linnemann}, \citenamefont
  {Zibold}, \citenamefont {Hume}, \citenamefont {Pezzè}, \citenamefont
  {Smerzi},\ and\ \citenamefont {Oberthaler}}]{Strobel_2014}%
  \BibitemOpen
  \bibfield  {author} {\bibinfo {author} {\bibfnamefont {H.}~\bibnamefont
  {Strobel}}, \bibinfo {author} {\bibfnamefont {W.}~\bibnamefont {Muessel}},
  \bibinfo {author} {\bibfnamefont {D.}~\bibnamefont {Linnemann}}, \bibinfo
  {author} {\bibfnamefont {T.}~\bibnamefont {Zibold}}, \bibinfo {author}
  {\bibfnamefont {D.~B.}\ \bibnamefont {Hume}}, \bibinfo {author}
  {\bibfnamefont {L.}~\bibnamefont {Pezzè}}, \bibinfo {author} {\bibfnamefont
  {A.}~\bibnamefont {Smerzi}},\ and\ \bibinfo {author} {\bibfnamefont {M.~K.}\
  \bibnamefont {Oberthaler}},\ }\bibfield  {title} {\bibinfo {title} {Fisher
  information and entanglement of non-gaussian spin states},\ }\href
  {https://doi.org/10.1126/science.1250147} {\bibfield  {journal} {\bibinfo
  {journal} {Science}\ }\textbf {\bibinfo {volume} {345}},\ \bibinfo {pages}
  {424–427} (\bibinfo {year} {2014})}\BibitemShut {NoStop}%
\bibitem [{\citenamefont {Bornet}\ \emph {et~al.}(2023)\citenamefont {Bornet},
  \citenamefont {Emperauger}, \citenamefont {Chen}, \citenamefont {Ye},
  \citenamefont {Block}, \citenamefont {Bintz}, \citenamefont {Boyd},
  \citenamefont {Barredo}, \citenamefont {Comparin}, \citenamefont {Mezzacapo},
  \citenamefont {Roscilde}, \citenamefont {Lahaye}, \citenamefont {Yao},\ and\
  \citenamefont {Browaeys}}]{Bornet_2023}%
  \BibitemOpen
  \bibfield  {author} {\bibinfo {author} {\bibfnamefont {G.}~\bibnamefont
  {Bornet}}, \bibinfo {author} {\bibfnamefont {G.}~\bibnamefont {Emperauger}},
  \bibinfo {author} {\bibfnamefont {C.}~\bibnamefont {Chen}}, \bibinfo {author}
  {\bibfnamefont {B.}~\bibnamefont {Ye}}, \bibinfo {author} {\bibfnamefont
  {M.}~\bibnamefont {Block}}, \bibinfo {author} {\bibfnamefont
  {M.}~\bibnamefont {Bintz}}, \bibinfo {author} {\bibfnamefont {J.~A.}\
  \bibnamefont {Boyd}}, \bibinfo {author} {\bibfnamefont {D.}~\bibnamefont
  {Barredo}}, \bibinfo {author} {\bibfnamefont {T.}~\bibnamefont {Comparin}},
  \bibinfo {author} {\bibfnamefont {F.}~\bibnamefont {Mezzacapo}}, \bibinfo
  {author} {\bibfnamefont {T.}~\bibnamefont {Roscilde}}, \bibinfo {author}
  {\bibfnamefont {T.}~\bibnamefont {Lahaye}}, \bibinfo {author} {\bibfnamefont
  {N.~Y.}\ \bibnamefont {Yao}},\ and\ \bibinfo {author} {\bibfnamefont
  {A.}~\bibnamefont {Browaeys}},\ }\bibfield  {title} {\bibinfo {title}
  {Scalable spin squeezing in a dipolar rydberg atom array},\ }\href
  {https://doi.org/10.1038/s41586-023-06414-9} {\bibfield  {journal} {\bibinfo
  {journal} {Nature}\ }\textbf {\bibinfo {volume} {621}},\ \bibinfo {pages}
  {728–733} (\bibinfo {year} {2023})}\BibitemShut {NoStop}%
\bibitem [{\citenamefont {Cheraghi}\ \emph {et~al.}(2022)\citenamefont
  {Cheraghi}, \citenamefont {Mahdavifar},\ and\ \citenamefont
  {Johannesson}}]{Cheraghi_2022}%
  \BibitemOpen
  \bibfield  {author} {\bibinfo {author} {\bibfnamefont {H.}~\bibnamefont
  {Cheraghi}}, \bibinfo {author} {\bibfnamefont {S.}~\bibnamefont
  {Mahdavifar}},\ and\ \bibinfo {author} {\bibfnamefont {H.}~\bibnamefont
  {Johannesson}},\ }\bibfield  {title} {\bibinfo {title} {Achieving
  spin-squeezed states by quench dynamics in a quantum chain},\ }\href
  {https://doi.org/10.1103/physrevb.105.024425} {\bibfield  {journal} {\bibinfo
   {journal} {Phys. Rev. B}\ }\textbf {\bibinfo {volume} {105}},\ \bibinfo
  {pages} {024425} (\bibinfo {year} {2022})}\BibitemShut {NoStop}%
\bibitem [{\citenamefont {Altafini}\ and\ \citenamefont
  {Ticozzi}(2012)}]{altafini_modeling_2012}%
  \BibitemOpen
  \bibfield  {author} {\bibinfo {author} {\bibfnamefont {C.}~\bibnamefont
  {Altafini}}\ and\ \bibinfo {author} {\bibfnamefont {F.}~\bibnamefont
  {Ticozzi}},\ }\bibfield  {title} {\bibinfo {title} {Modeling and {Control} of
  {Quantum} {Systems}: {An} {Introduction}},\ }\href
  {https://doi.org/10.1109/TAC.2012.2195830} {\bibfield  {journal} {\bibinfo
  {journal} {IEEE Trans. on Autom. Control}\ }\textbf {\bibinfo {volume}
  {57}},\ \bibinfo {pages} {1898} (\bibinfo {year} {2012})}\BibitemShut
  {NoStop}%
\bibitem [{\citenamefont {Glaser}\ \emph {et~al.}(2025)\citenamefont {Glaser},
  \citenamefont {Boscain}, \citenamefont {Calarco}, \citenamefont {Koch},
  \citenamefont {Köckenberger}, \citenamefont {Kosloff}, \citenamefont
  {Kuprov}, \citenamefont {Luy}, \citenamefont {Schirmer}, \citenamefont
  {Schulte-Herbrüggen}, \citenamefont {Sugny},\ and\ \citenamefont
  {Wilhelm}}]{glaserTrainingSchrodingersCat2015}%
  \BibitemOpen
  \bibfield  {author} {\bibinfo {author} {\bibfnamefont {S.~J.}\ \bibnamefont
  {Glaser}}, \bibinfo {author} {\bibfnamefont {U.}~\bibnamefont {Boscain}},
  \bibinfo {author} {\bibfnamefont {T.}~\bibnamefont {Calarco}}, \bibinfo
  {author} {\bibfnamefont {C.~P.}\ \bibnamefont {Koch}}, \bibinfo {author}
  {\bibfnamefont {W.}~\bibnamefont {Köckenberger}}, \bibinfo {author}
  {\bibfnamefont {R.}~\bibnamefont {Kosloff}}, \bibinfo {author} {\bibfnamefont
  {I.}~\bibnamefont {Kuprov}}, \bibinfo {author} {\bibfnamefont
  {B.}~\bibnamefont {Luy}}, \bibinfo {author} {\bibfnamefont {S.}~\bibnamefont
  {Schirmer}}, \bibinfo {author} {\bibfnamefont {T.}~\bibnamefont
  {Schulte-Herbrüggen}}, \bibinfo {author} {\bibfnamefont {D.}~\bibnamefont
  {Sugny}},\ and\ \bibinfo {author} {\bibfnamefont {F.~K.}\ \bibnamefont
  {Wilhelm}},\ }\bibfield  {title} {\bibinfo {title} {Training
  {{Schrödinger}}’s cat: Quantum optimal control},\ }\href
  {https://doi.org/10.1140/epjd/e2015-60464-1} {\bibfield  {journal} {\bibinfo
  {journal} {Eur. Phys. J. D}\ }\textbf {\bibinfo {volume} {69}},\ \bibinfo
  {pages} {279} (\bibinfo {year} {2025})}\BibitemShut {NoStop}%
\bibitem [{\citenamefont {Doria}\ \emph {et~al.}(2011)\citenamefont {Doria},
  \citenamefont {Calarco},\ and\ \citenamefont
  {Montangero}}]{doria_optimal_2011}%
  \BibitemOpen
  \bibfield  {author} {\bibinfo {author} {\bibfnamefont {P.}~\bibnamefont
  {Doria}}, \bibinfo {author} {\bibfnamefont {T.}~\bibnamefont {Calarco}},\
  and\ \bibinfo {author} {\bibfnamefont {S.}~\bibnamefont {Montangero}},\
  }\bibfield  {title} {\bibinfo {title} {Optimal {Control} {Technique} for
  {Many}-{Body} {Quantum} {Dynamics}},\ }\href
  {https://doi.org/10.1103/PhysRevLett.106.190501} {\bibfield  {journal}
  {\bibinfo  {journal} {Phys. Rev. Lett.}\ }\textbf {\bibinfo {volume} {106}},\
  \bibinfo {pages} {190501} (\bibinfo {year} {2011})}\BibitemShut {NoStop}%
\bibitem [{\citenamefont {Luchnikov}\ \emph {et~al.}(2024)\citenamefont
  {Luchnikov}, \citenamefont {Gavreev},\ and\ \citenamefont
  {Fedorov}}]{luchnikov_controlling_2024}%
  \BibitemOpen
  \bibfield  {author} {\bibinfo {author} {\bibfnamefont {I.~A.}\ \bibnamefont
  {Luchnikov}}, \bibinfo {author} {\bibfnamefont {M.~A.}\ \bibnamefont
  {Gavreev}},\ and\ \bibinfo {author} {\bibfnamefont {A.~K.}\ \bibnamefont
  {Fedorov}},\ }\bibfield  {title} {\bibinfo {title} {Controlling quantum
  many-body systems using reduced-order modeling},\ }\href
  {https://doi.org/10.1103/PhysRevResearch.6.013161} {\bibfield  {journal}
  {\bibinfo  {journal} {Phys. Rev. Res.}\ }\textbf {\bibinfo {volume} {6}},\
  \bibinfo {pages} {013161} (\bibinfo {year} {2024})}\BibitemShut {NoStop}%
\bibitem [{\citenamefont {Zeng}\ \emph {et~al.}(2025)\citenamefont {Zeng},
  \citenamefont {Giudici}, \citenamefont {Senoo}, \citenamefont
  {Baumg{\"a}rtner}, \citenamefont {Kaufman},\ and\ \citenamefont
  {Pichler}}]{zengAdiabaticEchoProtocols2025}%
  \BibitemOpen
  \bibfield  {author} {\bibinfo {author} {\bibfnamefont {Z.}~\bibnamefont
  {Zeng}}, \bibinfo {author} {\bibfnamefont {G.}~\bibnamefont {Giudici}},
  \bibinfo {author} {\bibfnamefont {A.}~\bibnamefont {Senoo}}, \bibinfo
  {author} {\bibfnamefont {A.}~\bibnamefont {Baumg{\"a}rtner}}, \bibinfo
  {author} {\bibfnamefont {A.~M.}\ \bibnamefont {Kaufman}},\ and\ \bibinfo
  {author} {\bibfnamefont {H.}~\bibnamefont {Pichler}},\ }\bibfield  {title}
  {\bibinfo {title} {Adiabatic echo protocols for robust quantum many-body
  state preparation},\ }\href@noop {} {\  (\bibinfo {year} {2025})},\ \Eprint
  {https://arxiv.org/abs/2506.12138} {arXiv:2506.12138 [quant-ph]} \BibitemShut
  {NoStop}%
\bibitem [{\citenamefont {Law}\ \emph {et~al.}(2001)\citenamefont {Law},
  \citenamefont {Ng},\ and\ \citenamefont {Leung}}]{Law_2001}%
  \BibitemOpen
  \bibfield  {author} {\bibinfo {author} {\bibfnamefont {C.~K.}\ \bibnamefont
  {Law}}, \bibinfo {author} {\bibfnamefont {H.~T.}\ \bibnamefont {Ng}},\ and\
  \bibinfo {author} {\bibfnamefont {P.~T.}\ \bibnamefont {Leung}},\ }\bibfield
  {title} {\bibinfo {title} {Coherent control of spin squeezing},\ }\href
  {https://doi.org/10.1103/physreva.63.055601} {\bibfield  {journal} {\bibinfo
  {journal} {Phys. Rev. A}\ }\textbf {\bibinfo {volume} {63}},\ \bibinfo
  {pages} {055601} (\bibinfo {year} {2001})}\BibitemShut {NoStop}%
\bibitem [{\citenamefont {Shen}\ and\ \citenamefont
  {Duan}(2013)}]{shenEfficientSpinSqueezing2013}%
  \BibitemOpen
  \bibfield  {author} {\bibinfo {author} {\bibfnamefont {C.}~\bibnamefont
  {Shen}}\ and\ \bibinfo {author} {\bibfnamefont {L.-M.}\ \bibnamefont
  {Duan}},\ }\bibfield  {title} {\bibinfo {title} {Efficient spin squeezing
  with optimized pulse sequences},\ }\href
  {https://doi.org/10.1103/PhysRevA.87.051801} {\bibfield  {journal} {\bibinfo
  {journal} {Phys. Rev. A}\ }\textbf {\bibinfo {volume} {87}},\ \bibinfo
  {pages} {051801} (\bibinfo {year} {2013})}\BibitemShut {NoStop}%
\bibitem [{\citenamefont {Pichler}\ \emph {et~al.}(2016)\citenamefont
  {Pichler}, \citenamefont {Caneva}, \citenamefont {Montangero}, \citenamefont
  {Lukin},\ and\ \citenamefont {Calarco}}]{Pichler_2016}%
  \BibitemOpen
  \bibfield  {author} {\bibinfo {author} {\bibfnamefont {T.}~\bibnamefont
  {Pichler}}, \bibinfo {author} {\bibfnamefont {T.}~\bibnamefont {Caneva}},
  \bibinfo {author} {\bibfnamefont {S.}~\bibnamefont {Montangero}}, \bibinfo
  {author} {\bibfnamefont {M.~D.}\ \bibnamefont {Lukin}},\ and\ \bibinfo
  {author} {\bibfnamefont {T.}~\bibnamefont {Calarco}},\ }\bibfield  {title}
  {\bibinfo {title} {Noise-resistant optimal spin squeezing via quantum
  control},\ }\href {https://doi.org/10.1103/physreva.93.013851} {\bibfield
  {journal} {\bibinfo  {journal} {Phys. Rev. A}\ }\textbf {\bibinfo {volume}
  {93}},\ \bibinfo {pages} {013851} (\bibinfo {year} {2016})}\BibitemShut
  {NoStop}%
\bibitem [{\citenamefont {Haine}\ and\ \citenamefont
  {Hope}(2020)}]{haineMachineDesignedSensorMake2020}%
  \BibitemOpen
  \bibfield  {author} {\bibinfo {author} {\bibfnamefont {S.~A.}\ \bibnamefont
  {Haine}}\ and\ \bibinfo {author} {\bibfnamefont {J.~J.}\ \bibnamefont
  {Hope}},\ }\bibfield  {title} {\bibinfo {title} {Machine-{{Designed Sensor}}
  to {{Make Optimal Use}} of {{Entanglement-Generating Dynamics}} for {{Quantum
  Sensing}}},\ }\href {https://doi.org/10.1103/PhysRevLett.124.060402}
  {\bibfield  {journal} {\bibinfo  {journal} {Phys. Rev. Lett.}\ }\textbf
  {\bibinfo {volume} {124}},\ \bibinfo {pages} {060402} (\bibinfo {year}
  {2020})}\BibitemShut {NoStop}%
\bibitem [{\citenamefont {Zhao}\ \emph
  {et~al.}(2024{\natexlab{a}})\citenamefont {Zhao}, \citenamefont {Zhao},
  \citenamefont {Li}, \citenamefont {Li}, \citenamefont {Liu}, \citenamefont
  {Guo},\ and\ \citenamefont {Yi}}]{zhao_strategy_2024}%
  \BibitemOpen
  \bibfield  {author} {\bibinfo {author} {\bibfnamefont {X.~L.}\ \bibnamefont
  {Zhao}}, \bibinfo {author} {\bibfnamefont {Y.~M.}\ \bibnamefont {Zhao}},
  \bibinfo {author} {\bibfnamefont {M.}~\bibnamefont {Li}}, \bibinfo {author}
  {\bibfnamefont {T.~T.}\ \bibnamefont {Li}}, \bibinfo {author} {\bibfnamefont
  {Q.}~\bibnamefont {Liu}}, \bibinfo {author} {\bibfnamefont {S.}~\bibnamefont
  {Guo}},\ and\ \bibinfo {author} {\bibfnamefont {X.~X.}\ \bibnamefont {Yi}},\
  }\bibfield  {title} {\bibinfo {title} {A {Strategy} for {Preparing} {Quantum}
  {Squeezed} {States} {Using} {Reinforcement} {Learning}},\ }\href
  {http://arxiv.org/abs/2401.16320} {\  (\bibinfo {year}
  {2024}{\natexlab{a}})},\ \Eprint {https://arxiv.org/abs/2401.16320}
  {arXiv:2401.16320} \BibitemShut {NoStop}%
\bibitem [{\citenamefont {Zhao}\ \emph
  {et~al.}(2024{\natexlab{b}})\citenamefont {Zhao}, \citenamefont {Chen},
  \citenamefont {Wang}, \citenamefont {Ma},\ and\ \citenamefont
  {Zhao}}]{zhao_preparing_2024}%
  \BibitemOpen
  \bibfield  {author} {\bibinfo {author} {\bibfnamefont {Y.}~\bibnamefont
  {Zhao}}, \bibinfo {author} {\bibfnamefont {L.}~\bibnamefont {Chen}}, \bibinfo
  {author} {\bibfnamefont {Y.}~\bibnamefont {Wang}}, \bibinfo {author}
  {\bibfnamefont {H.}~\bibnamefont {Ma}},\ and\ \bibinfo {author}
  {\bibfnamefont {X.}~\bibnamefont {Zhao}},\ }\bibfield  {title} {\bibinfo
  {title} {Preparing {Spin} {Squeezed} {States} via {Adaptive} {Genetic}
  {Algorithm}},\ }\href {https://arxiv.org/abs/2410.15375} {\  (\bibinfo {year}
  {2024}{\natexlab{b}})},\ \Eprint {https://arxiv.org/abs/2410.15375}
  {arXiv:2410.15375} \BibitemShut {NoStop}%
\bibitem [{\citenamefont {Carrera}\ \emph {et~al.}(2025)\citenamefont
  {Carrera}, \citenamefont {Zhang}, \citenamefont {Bancal},\ and\ \citenamefont
  {Sangouard}}]{carrera_testing_2025}%
  \BibitemOpen
  \bibfield  {author} {\bibinfo {author} {\bibfnamefont {E.}~\bibnamefont
  {Carrera}}, \bibinfo {author} {\bibfnamefont {Y.}~\bibnamefont {Zhang}},
  \bibinfo {author} {\bibfnamefont {J.-D.}\ \bibnamefont {Bancal}},\ and\
  \bibinfo {author} {\bibfnamefont {N.}~\bibnamefont {Sangouard}},\ }\bibfield
  {title} {\bibinfo {title} {Testing the {Wineland} {Criterion} with {Finite}
  {Statistics}},\ }\href {https://doi.org/10.1103/5svm-skxk} {\bibfield
  {journal} {\bibinfo  {journal} {Phys. Rev. Lett.}\ }\textbf {\bibinfo
  {volume} {134}},\ \bibinfo {pages} {220802} (\bibinfo {year}
  {2025})}\BibitemShut {NoStop}%
\bibitem [{\citenamefont {Saffman}\ \emph {et~al.}(2010)\citenamefont
  {Saffman}, \citenamefont {Walker},\ and\ \citenamefont
  {Mølmer}}]{saffman_quantum_2010}%
  \BibitemOpen
  \bibfield  {author} {\bibinfo {author} {\bibfnamefont {M.}~\bibnamefont
  {Saffman}}, \bibinfo {author} {\bibfnamefont {T.~G.}\ \bibnamefont
  {Walker}},\ and\ \bibinfo {author} {\bibfnamefont {K.}~\bibnamefont
  {Mølmer}},\ }\bibfield  {title} {\bibinfo {title} {Quantum information with
  {Rydberg} atoms},\ }\href {https://doi.org/10.1103/RevModPhys.82.2313}
  {\bibfield  {journal} {\bibinfo  {journal} {Rev. Mod. Phys.}\ }\textbf
  {\bibinfo {volume} {82}},\ \bibinfo {pages} {2313} (\bibinfo {year}
  {2010})}\BibitemShut {NoStop}%
\bibitem [{\citenamefont {Browaeys}\ and\ \citenamefont
  {Lahaye}(2020)}]{browaeys_many-body_2020}%
  \BibitemOpen
  \bibfield  {author} {\bibinfo {author} {\bibfnamefont {A.}~\bibnamefont
  {Browaeys}}\ and\ \bibinfo {author} {\bibfnamefont {T.}~\bibnamefont
  {Lahaye}},\ }\bibfield  {title} {\bibinfo {title} {Many-body physics with
  individually controlled {Rydberg} atoms},\ }\href
  {https://doi.org/10.1038/s41567-019-0733-z} {\bibfield  {journal} {\bibinfo
  {journal} {Nat. Phys.}\ }\textbf {\bibinfo {volume} {16}},\ \bibinfo {pages}
  {132} (\bibinfo {year} {2020})}\BibitemShut {NoStop}%
\bibitem [{\citenamefont {Henriet}\ \emph {et~al.}(2020)\citenamefont
  {Henriet}, \citenamefont {Beguin}, \citenamefont {Signoles}, \citenamefont
  {Lahaye}, \citenamefont {Browaeys}, \citenamefont {Reymond},\ and\
  \citenamefont {Jurczak}}]{Henriet_2020}%
  \BibitemOpen
  \bibfield  {author} {\bibinfo {author} {\bibfnamefont {L.}~\bibnamefont
  {Henriet}}, \bibinfo {author} {\bibfnamefont {L.}~\bibnamefont {Beguin}},
  \bibinfo {author} {\bibfnamefont {A.}~\bibnamefont {Signoles}}, \bibinfo
  {author} {\bibfnamefont {T.}~\bibnamefont {Lahaye}}, \bibinfo {author}
  {\bibfnamefont {A.}~\bibnamefont {Browaeys}}, \bibinfo {author}
  {\bibfnamefont {G.-O.}\ \bibnamefont {Reymond}},\ and\ \bibinfo {author}
  {\bibfnamefont {C.}~\bibnamefont {Jurczak}},\ }\bibfield  {title} {\bibinfo
  {title} {Quantum computing with neutral atoms},\ }\href
  {https://doi.org/10.22331/q-2020-09-21-327} {\bibfield  {journal} {\bibinfo
  {journal} {Quantum}\ }\textbf {\bibinfo {volume} {4}},\ \bibinfo {pages}
  {327} (\bibinfo {year} {2020})}\BibitemShut {NoStop}%
\bibitem [{\citenamefont {Béguin}\ \emph {et~al.}(2013)\citenamefont
  {Béguin}, \citenamefont {Vernier}, \citenamefont {Chicireanu}, \citenamefont
  {Lahaye},\ and\ \citenamefont {Browaeys}}]{beguin_direct_2013}%
  \BibitemOpen
  \bibfield  {author} {\bibinfo {author} {\bibfnamefont {L.}~\bibnamefont
  {Béguin}}, \bibinfo {author} {\bibfnamefont {A.}~\bibnamefont {Vernier}},
  \bibinfo {author} {\bibfnamefont {R.}~\bibnamefont {Chicireanu}}, \bibinfo
  {author} {\bibfnamefont {T.}~\bibnamefont {Lahaye}},\ and\ \bibinfo {author}
  {\bibfnamefont {A.}~\bibnamefont {Browaeys}},\ }\bibfield  {title} {\bibinfo
  {title} {Direct {Measurement} of the van der {Waals} {Interaction} between
  {Two} {Rydberg} {Atoms}},\ }\href
  {https://doi.org/10.1103/PhysRevLett.110.263201} {\bibfield  {journal}
  {\bibinfo  {journal} {Phys. Rev. Lett.}\ }\textbf {\bibinfo {volume} {110}},\
  \bibinfo {pages} {263201} (\bibinfo {year} {2013})}\BibitemShut {NoStop}%
\bibitem [{\citenamefont {Wurtz}\ \emph {et~al.}(2023)\citenamefont {Wurtz},
  \citenamefont {Bylinskii}, \citenamefont {Braverman}, \citenamefont
  {Amato-Grill}, \citenamefont {Cantu}, \citenamefont {Huber}, \citenamefont
  {Lukin}, \citenamefont {Liu}, \citenamefont {Weinberg}, \citenamefont {Long},
  \citenamefont {Wang}, \citenamefont {Gemelke},\ and\ \citenamefont
  {Keesling}}]{wurtz_aquila_2023}%
  \BibitemOpen
  \bibfield  {author} {\bibinfo {author} {\bibfnamefont {J.}~\bibnamefont
  {Wurtz}}, \bibinfo {author} {\bibfnamefont {A.}~\bibnamefont {Bylinskii}},
  \bibinfo {author} {\bibfnamefont {B.}~\bibnamefont {Braverman}}, \bibinfo
  {author} {\bibfnamefont {J.}~\bibnamefont {Amato-Grill}}, \bibinfo {author}
  {\bibfnamefont {S.~H.}\ \bibnamefont {Cantu}}, \bibinfo {author}
  {\bibfnamefont {F.}~\bibnamefont {Huber}}, \bibinfo {author} {\bibfnamefont
  {A.}~\bibnamefont {Lukin}}, \bibinfo {author} {\bibfnamefont
  {F.}~\bibnamefont {Liu}}, \bibinfo {author} {\bibfnamefont {P.}~\bibnamefont
  {Weinberg}}, \bibinfo {author} {\bibfnamefont {J.}~\bibnamefont {Long}},
  \bibinfo {author} {\bibfnamefont {S.-T.}\ \bibnamefont {Wang}}, \bibinfo
  {author} {\bibfnamefont {N.}~\bibnamefont {Gemelke}},\ and\ \bibinfo {author}
  {\bibfnamefont {A.}~\bibnamefont {Keesling}},\ }\bibfield  {title} {\bibinfo
  {title} {Aquila: {QuEra}'s 256-qubit neutral-atom quantum computer},\ }\href
  {http://arxiv.org/abs/2306.11727} {\  (\bibinfo {year} {2023})},\ \Eprint
  {https://arxiv.org/abs/2306.11727} {arXiv:2306.11727} \BibitemShut {NoStop}%
\bibitem [{\citenamefont {de~Léséleuc}\ \emph {et~al.}(2018)\citenamefont
  {de~Léséleuc}, \citenamefont {Barredo}, \citenamefont {Lienhard},
  \citenamefont {Browaeys},\ and\ \citenamefont
  {Lahaye}}]{leseleuc_analysis_2018}%
  \BibitemOpen
  \bibfield  {author} {\bibinfo {author} {\bibfnamefont {S.}~\bibnamefont
  {de~Léséleuc}}, \bibinfo {author} {\bibfnamefont {D.}~\bibnamefont
  {Barredo}}, \bibinfo {author} {\bibfnamefont {V.}~\bibnamefont {Lienhard}},
  \bibinfo {author} {\bibfnamefont {A.}~\bibnamefont {Browaeys}},\ and\
  \bibinfo {author} {\bibfnamefont {T.}~\bibnamefont {Lahaye}},\ }\bibfield
  {title} {\bibinfo {title} {Analysis of imperfections in the coherent optical
  excitation of single atoms to {{Rydberg}} states},\ }\href
  {https://doi.org/10.1103/PhysRevA.97.053803} {\bibfield  {journal} {\bibinfo
  {journal} {Phys. Rev. A}\ }\textbf {\bibinfo {volume} {97}},\ \bibinfo
  {pages} {053803} (\bibinfo {year} {2018})}\BibitemShut {NoStop}%
\bibitem [{\citenamefont {Baydin}\ \emph {et~al.}(2018)\citenamefont {Baydin},
  \citenamefont {Pearlmutter}, \citenamefont {Radul},\ and\ \citenamefont
  {Siskind}}]{baydin2018automaticdifferentiationmachinelearning}%
  \BibitemOpen
  \bibfield  {author} {\bibinfo {author} {\bibfnamefont {A.~G.}\ \bibnamefont
  {Baydin}}, \bibinfo {author} {\bibfnamefont {B.~A.}\ \bibnamefont
  {Pearlmutter}}, \bibinfo {author} {\bibfnamefont {A.~A.}\ \bibnamefont
  {Radul}},\ and\ \bibinfo {author} {\bibfnamefont {J.~M.}\ \bibnamefont
  {Siskind}},\ }\bibfield  {title} {\bibinfo {title} {Automatic differentiation
  in machine learning: A survey},\ }\href
  {https://www.jmlr.org/papers/v18/17-468.html} {\bibfield  {journal} {\bibinfo
   {journal} {J. Mach. Learn. Res.}\ }\textbf {\bibinfo {volume} {18}},\
  \bibinfo {pages} {1} (\bibinfo {year} {2018})}\BibitemShut {NoStop}%
\bibitem [{\citenamefont {Khaneja}\ \emph {et~al.}(2005)\citenamefont
  {Khaneja}, \citenamefont {Reiss}, \citenamefont {Kehlet}, \citenamefont
  {Schulte-Herbr\"{u}ggen},\ and\ \citenamefont {Glaser}}]{Khaneja2005}%
  \BibitemOpen
  \bibfield  {author} {\bibinfo {author} {\bibfnamefont {N.}~\bibnamefont
  {Khaneja}}, \bibinfo {author} {\bibfnamefont {T.}~\bibnamefont {Reiss}},
  \bibinfo {author} {\bibfnamefont {C.}~\bibnamefont {Kehlet}}, \bibinfo
  {author} {\bibfnamefont {T.}~\bibnamefont {Schulte-Herbr\"{u}ggen}},\ and\
  \bibinfo {author} {\bibfnamefont {S.~J.}\ \bibnamefont {Glaser}},\ }\bibfield
   {title} {\bibinfo {title} {Optimal control of coupled spin dynamics: design
  of nmr pulse sequences by gradient ascent algorithms},\ }\href
  {https://doi.org/10.1016/j.jmr.2004.11.004} {\bibfield  {journal} {\bibinfo
  {journal} {J. Mag. Res.}\ }\textbf {\bibinfo {volume} {172}},\ \bibinfo
  {pages} {296–305} (\bibinfo {year} {2005})}\BibitemShut {NoStop}%
\bibitem [{\citenamefont {Saywell}\ \emph {et~al.}(2018)\citenamefont
  {Saywell}, \citenamefont {Kuprov}, \citenamefont {Goodwin}, \citenamefont
  {Carey},\ and\ \citenamefont {Freegarde}}]{Saywell2018}%
  \BibitemOpen
  \bibfield  {author} {\bibinfo {author} {\bibfnamefont {J.~C.}\ \bibnamefont
  {Saywell}}, \bibinfo {author} {\bibfnamefont {I.}~\bibnamefont {Kuprov}},
  \bibinfo {author} {\bibfnamefont {D.}~\bibnamefont {Goodwin}}, \bibinfo
  {author} {\bibfnamefont {M.}~\bibnamefont {Carey}},\ and\ \bibinfo {author}
  {\bibfnamefont {T.}~\bibnamefont {Freegarde}},\ }\bibfield  {title} {\bibinfo
  {title} {Optimal control of mirror pulses for cold-atom interferometry},\
  }\href {https://doi.org/10.1103/physreva.98.023625} {\bibfield  {journal}
  {\bibinfo  {journal} {Phys. Rev. A}\ }\textbf {\bibinfo {volume} {98}},\
  \bibinfo {pages} {023625} (\bibinfo {year} {2018})}\BibitemShut {NoStop}%
\bibitem [{\citenamefont {Bradbury}\ \emph {et~al.}(2018)\citenamefont
  {Bradbury}, \citenamefont {Frostig}, \citenamefont {Hawkins}, \citenamefont
  {Johnson}, \citenamefont {Leary}, \citenamefont {Maclaurin}, \citenamefont
  {Necula}, \citenamefont {Paszke}, \citenamefont {Vander{P}las}, \citenamefont
  {Wanderman-{M}ilne},\ and\ \citenamefont {Zhang}}]{jax2018github}%
  \BibitemOpen
  \bibfield  {author} {\bibinfo {author} {\bibfnamefont {J.}~\bibnamefont
  {Bradbury}}, \bibinfo {author} {\bibfnamefont {R.}~\bibnamefont {Frostig}},
  \bibinfo {author} {\bibfnamefont {P.}~\bibnamefont {Hawkins}}, \bibinfo
  {author} {\bibfnamefont {M.~J.}\ \bibnamefont {Johnson}}, \bibinfo {author}
  {\bibfnamefont {C.}~\bibnamefont {Leary}}, \bibinfo {author} {\bibfnamefont
  {D.}~\bibnamefont {Maclaurin}}, \bibinfo {author} {\bibfnamefont
  {G.}~\bibnamefont {Necula}}, \bibinfo {author} {\bibfnamefont
  {A.}~\bibnamefont {Paszke}}, \bibinfo {author} {\bibfnamefont
  {J.}~\bibnamefont {Vander{P}las}}, \bibinfo {author} {\bibfnamefont
  {S.}~\bibnamefont {Wanderman-{M}ilne}},\ and\ \bibinfo {author}
  {\bibfnamefont {Q.}~\bibnamefont {Zhang}},\ }\bibfield  {title} {\bibinfo
  {title} {{JAX}: composable transformations of {P}ython+{N}um{P}y programs},\
  }\href {http://github.com/jax-ml/jax} {\  (\bibinfo {year} {2018})},\
  \bibinfo {note}
  {\href{https://github.com/jax-ml/jax}{github.com/jax-ml/jax}}\BibitemShut
  {NoStop}%
\bibitem [{\citenamefont {Husimi}(1940)}]{husimi_formal_1940}%
  \BibitemOpen
  \bibfield  {author} {\bibinfo {author} {\bibfnamefont {K.}~\bibnamefont
  {Husimi}},\ }\bibfield  {title} {\bibinfo {title} {Some {Formal} {Properties}
  of the {Density} {Matrix}},\ }\href
  {https://doi.org/10.11429/ppmsj1919.22.4_264} {\bibfield  {journal} {\bibinfo
   {journal} {Proc. Phys.-Math. Soc. Jpn. 3rd Series}\ }\textbf {\bibinfo
  {volume} {22}},\ \bibinfo {pages} {264} (\bibinfo {year} {1940})}\BibitemShut
  {NoStop}%
\bibitem [{\citenamefont {Guilmin}\ \emph {et~al.}(2025)\citenamefont
  {Guilmin}, \citenamefont {Bocquet}, \citenamefont {Genois}, \citenamefont
  {Weiss},\ and\ \citenamefont {Gautier}}]{guilmin2025dynamiqs}%
  \BibitemOpen
  \bibfield  {author} {\bibinfo {author} {\bibfnamefont {P.}~\bibnamefont
  {Guilmin}}, \bibinfo {author} {\bibfnamefont {A.}~\bibnamefont {Bocquet}},
  \bibinfo {author} {\bibfnamefont {{\'{E}}.}~\bibnamefont {Genois}}, \bibinfo
  {author} {\bibfnamefont {D.}~\bibnamefont {Weiss}},\ and\ \bibinfo {author}
  {\bibfnamefont {R.}~\bibnamefont {Gautier}},\ }\bibfield  {title} {\bibinfo
  {title} {Dynamiqs: an open-source python library for gpu-accelerated and
  differentiable simulation of quantum systems}} (\bibinfo {year} {2025}),\
  \bibinfo {note}
  {\href{https://github.com/dynamiqs/dynamiqs}{github.com/dynamiqs/dynamiqs}}\BibitemShut
  {NoStop}%
\bibitem [{\citenamefont {Vidal}\ \emph {et~al.}(2004)\citenamefont {Vidal},
  \citenamefont {Palacios},\ and\ \citenamefont
  {Mosseri}}]{vidalEntanglementSecondorderQuantum2004}%
  \BibitemOpen
  \bibfield  {author} {\bibinfo {author} {\bibfnamefont {J.}~\bibnamefont
  {Vidal}}, \bibinfo {author} {\bibfnamefont {G.}~\bibnamefont {Palacios}},\
  and\ \bibinfo {author} {\bibfnamefont {R.}~\bibnamefont {Mosseri}},\
  }\bibfield  {title} {\bibinfo {title} {Entanglement in a second-order quantum
  phase transition},\ }\href {https://doi.org/10.1103/PhysRevA.69.022107}
  {\bibfield  {journal} {\bibinfo  {journal} {Phys. Rev. A}\ }\textbf {\bibinfo
  {volume} {69}},\ \bibinfo {pages} {022107} (\bibinfo {year}
  {2004})}\BibitemShut {NoStop}%
\bibitem [{\citenamefont {Fr{\'e}rot}\ and\ \citenamefont
  {Roscilde}(2018)}]{frerotQuantumCriticalMetrology2018}%
  \BibitemOpen
  \bibfield  {author} {\bibinfo {author} {\bibfnamefont {I.}~\bibnamefont
  {Fr{\'e}rot}}\ and\ \bibinfo {author} {\bibfnamefont {T.}~\bibnamefont
  {Roscilde}},\ }\bibfield  {title} {\bibinfo {title} {Quantum {{Critical
  Metrology}}},\ }\href {https://doi.org/10.1103/PhysRevLett.121.020402}
  {\bibfield  {journal} {\bibinfo  {journal} {Phys. Rev. Lett.}\ }\textbf
  {\bibinfo {volume} {121}},\ \bibinfo {pages} {020402} (\bibinfo {year}
  {2018})}\BibitemShut {NoStop}%
\bibitem [{\citenamefont {Verresen}\ \emph {et~al.}(2021)\citenamefont
  {Verresen}, \citenamefont {Lukin},\ and\ \citenamefont
  {Vishwanath}}]{verresen_prediction_2021}%
  \BibitemOpen
  \bibfield  {author} {\bibinfo {author} {\bibfnamefont {R.}~\bibnamefont
  {Verresen}}, \bibinfo {author} {\bibfnamefont {M.~D.}\ \bibnamefont
  {Lukin}},\ and\ \bibinfo {author} {\bibfnamefont {A.}~\bibnamefont
  {Vishwanath}},\ }\bibfield  {title} {\bibinfo {title} {Prediction of {Toric}
  {Code} {Topological} {Order} from {Rydberg} {Blockade}},\ }\href
  {https://doi.org/10.1103/PhysRevX.11.031005} {\bibfield  {journal} {\bibinfo
  {journal} {Phys. Rev. X}\ }\textbf {\bibinfo {volume} {11}},\ \bibinfo
  {pages} {031005} (\bibinfo {year} {2021})}\BibitemShut {NoStop}%
\bibitem [{\citenamefont {Samajdar}\ \emph {et~al.}(2021)\citenamefont
  {Samajdar}, \citenamefont {Ho}, \citenamefont {Pichler}, \citenamefont
  {Lukin},\ and\ \citenamefont {Sachdev}}]{samajdar_quantum_2021}%
  \BibitemOpen
  \bibfield  {author} {\bibinfo {author} {\bibfnamefont {R.}~\bibnamefont
  {Samajdar}}, \bibinfo {author} {\bibfnamefont {W.~W.}\ \bibnamefont {Ho}},
  \bibinfo {author} {\bibfnamefont {H.}~\bibnamefont {Pichler}}, \bibinfo
  {author} {\bibfnamefont {M.~D.}\ \bibnamefont {Lukin}},\ and\ \bibinfo
  {author} {\bibfnamefont {S.}~\bibnamefont {Sachdev}},\ }\bibfield  {title}
  {\bibinfo {title} {Quantum phases of {Rydberg} atoms on a kagome lattice},\
  }\href {https://doi.org/10.1073/pnas.2015785118} {\bibfield  {journal}
  {\bibinfo  {journal} {PNAS}\ }\textbf {\bibinfo {volume} {118}},\ \bibinfo
  {pages} {e2015785118} (\bibinfo {year} {2021})}\BibitemShut {NoStop}%
\bibitem [{edi()}]{edison_github}%
  \BibitemOpen
  \href@noop {} {}\bibinfo {howpublished}
  {\url{https://github.com/edison651/squeezing_by_control}}\BibitemShut
  {NoStop}%
\bibitem [{\citenamefont {Kingma}\ and\ \citenamefont
  {Ba}(2017)}]{kingma_adam_2017}%
  \BibitemOpen
  \bibfield  {author} {\bibinfo {author} {\bibfnamefont {D.~P.}\ \bibnamefont
  {Kingma}}\ and\ \bibinfo {author} {\bibfnamefont {J.}~\bibnamefont {Ba}},\
  }\bibfield  {title} {\bibinfo {title} {Adam: {A} {Method} for {Stochastic}
  {Optimization}},\ }\href {http://arxiv.org/abs/1412.6980} {\  (\bibinfo
  {year} {2017})},\ \Eprint {https://arxiv.org/abs/1412.6980} {arXiv:1412.6980}
  \BibitemShut {NoStop}%
\bibitem [{\citenamefont {Tóth}\ and\ \citenamefont
  {Gühne}(2005)}]{toth_detecting_2005}%
  \BibitemOpen
  \bibfield  {author} {\bibinfo {author} {\bibfnamefont {G.}~\bibnamefont
  {Tóth}}\ and\ \bibinfo {author} {\bibfnamefont {O.}~\bibnamefont {Gühne}},\
  }\bibfield  {title} {\bibinfo {title} {Detecting {Genuine} {Multipartite}
  {Entanglement} with {Two} {Local} {Measurements}},\ }\href
  {https://doi.org/10.1103/PhysRevLett.94.060501} {\bibfield  {journal}
  {\bibinfo  {journal} {Phys. Rev. Lett.}\ }\textbf {\bibinfo {volume} {94}},\
  \bibinfo {pages} {060501} (\bibinfo {year} {2005})}\BibitemShut {NoStop}%
\bibitem [{\citenamefont {Sackett}\ \emph {et~al.}(2000)\citenamefont
  {Sackett}, \citenamefont {Kielpinski}, \citenamefont {King}, \citenamefont
  {Langer}, \citenamefont {Meyer}, \citenamefont {Myatt}, \citenamefont {Rowe},
  \citenamefont {Turchette}, \citenamefont {Itano}, \citenamefont {Wineland},\
  and\ \citenamefont {Monroe}}]{Sackett2000}%
  \BibitemOpen
  \bibfield  {author} {\bibinfo {author} {\bibfnamefont {C.~A.}\ \bibnamefont
  {Sackett}}, \bibinfo {author} {\bibfnamefont {D.}~\bibnamefont {Kielpinski}},
  \bibinfo {author} {\bibfnamefont {B.~E.}\ \bibnamefont {King}}, \bibinfo
  {author} {\bibfnamefont {C.}~\bibnamefont {Langer}}, \bibinfo {author}
  {\bibfnamefont {V.}~\bibnamefont {Meyer}}, \bibinfo {author} {\bibfnamefont
  {C.~J.}\ \bibnamefont {Myatt}}, \bibinfo {author} {\bibfnamefont
  {M.}~\bibnamefont {Rowe}}, \bibinfo {author} {\bibfnamefont {Q.~A.}\
  \bibnamefont {Turchette}}, \bibinfo {author} {\bibfnamefont {W.~M.}\
  \bibnamefont {Itano}}, \bibinfo {author} {\bibfnamefont {D.~J.}\ \bibnamefont
  {Wineland}},\ and\ \bibinfo {author} {\bibfnamefont {C.}~\bibnamefont
  {Monroe}},\ }\bibfield  {title} {\bibinfo {title} {Experimental entanglement
  of four particles},\ }\href {https://doi.org/10.1038/35005011} {\bibfield
  {journal} {\bibinfo  {journal} {Nature}\ }\textbf {\bibinfo {volume} {404}},\
  \bibinfo {pages} {256–259} (\bibinfo {year} {2000})}\BibitemShut {NoStop}%
\end{thebibliography}%

\end{document}